\documentclass{article}

\usepackage{PRIMEarxiv}

\usepackage{subfigure}
\usepackage{tikz}
\usepackage{multicol}
\usepackage{multirow}
\usepackage{makecell}
\usepackage{tabularx}
\usepackage{enumitem}
\usepackage[ruled,linesnumbered]{algorithm2e}
\usepackage{url}
\usepackage{graphicx}

\usepackage{amssymb}

\usepackage{titlesec}

\title{DFlow: Efficient Dataflow-based Invocation Workflow Execution for Function-as-a-Service}

\author{
  Xiaoxiang Shi  \\
  Shanghai Jiao Tong University  \\
  \texttt{lambda7shi@sjtu.edu.cn} \\
  \And 
  Chao Li \\
   Shanghai Jiao Tong University  \\
    \texttt{lichao@cs.sjtu.edu.cn} \\
  \And 
  Zijun Li \\ 
 Shanghai Jiao Tong University \\
 \texttt{lzjzx1122@sjtu.edu.cn}
 \And 
  Zihan Liu \\
   Shanghai Jiao Tong University  \\
  \texttt{altair.liu@sjtu.edu.cn} \\
  \And 
    Dianmo Sheng \\
 University of Science and Technolofy \\
  \texttt{dmsheng@mail.ustc.edu.cn} \\
  \And 
  Quan Chen \\
Shanghai Jiao Tong University  \\
\texttt{chen-quan@sjtu.edu.cn} \\
    \And 
    Jingwen Leng  \\
  Shanghai Jiao Tong University  \\
  \texttt{leng-jw@sjtu.edu.cn} \\
  \And 
  Minyi Guo \\
    Shanghai Jiao Tong University  \\
    \texttt{guo-my@cs.sjtu.edu.cn} \\
}

\begin{document}

\maketitle

\begin{abstract}
The Serverless Computing is becoming increasingly popular due to its ease of use and fine-grained billing. These features make it
appealing for stateful application or serverless workflow. However, current serverless workflow systems utilize a controlflow-based invocation pattern to invoke functions. In this execution pattern, the function invocation depends on the state of the function. A function can only begin executing once all its precursor functions have completed. As a result, this pattern may potentially lead to longer end-to-end execution time.

We design and implement the \textit{DFlow}, a novel dataflow-based serverless workflow system that achieves high performance for serverless workflow. DFlow introduces a distributed scheduler (\textit{DScheduler}) by using the dataflow-based invocation pattern to invoke functions. In this pattern, the function invocation depends on the data dependency between functions. The function can start to execute even its precursor functions are still running. DFlow further features a distributed store (\textit{DStore}) that utilizes effective fine-grained optimization techniques to eliminate function interaction, thereby enabling efficient data exchange. With the support of DScheduler and DStore, DFlow can achieving an average improvement of 60\% over CFlow, 40\% over FaaSFlow, 25\% over FaasFlowRedis, and 40\% over KNIX on \textit{99\%-ile latency} respectively. Further, it can improve network bandwidth utilization by 2$\times$-4$\times$ over CFlow and 1.5$\times$-3$\times$ over FaaSFlow, FaaSFlowRedis and KNIX, respectively. DFlow effectively reduces the cold startup latency, achieving an average improvement of 5.6$\times$ over CFlow and 1.1$\times$ over FaaSFlow.
\end{abstract}


\section{Introduction}

Serverless computing, also known as Function-as-a-Service (FaaS), has become increasingly popular in recent years. In serverless computing, users can employ high-level programming languages, such as Python, JavaScript, and Go, among others, to develop their functions. They then upload the code to the serverless platform, which takes on the responsibility of setting up the environment, executing the functions, and managing resources. This approach allows users to focus on their code and business logic while the platform handles underlying infrastructure concerns. Due to its usability, auto-scaling, and fine-grained cost model, serverless computing becomes attractive to emerging applications~\cite{SIGMOD20-survey} like machine learning and data analytics, forming stateful serverless computing or serverless workflow. A serverless workflow contains many functions that can be represented by a Directed Acyclic Graph (DAG). Within the Directed Acyclic Graph (DAG), each node signifies a function, and each edge represents a data dependency between a source function and a destination function. 



Many recent studies like FaaSFlow, SAND, Cloudburst and others ~\cite{FaaSFlow,cloudburst,Faastlane-ATC21,SAND-ATC18,Jiffy-EuroSys22,SIGMOD20-lambda,SIGMOD-Serverless-ML} all focus on serverless computing or serverless workflow. Several cloud platforms like AWS step Functions~\cite{AWS-step-functions, GoogleCloudFunctions, Alibaba-serverless-workflow, AzureFunctions} all add support for serverless workflow. A few open source systems like Fission~\cite{Fission, FN, OpenWhisk} also support serverless workflow. By analyzing these works, we observe that most serverless workflow systems follow a common invocation pattern, which we call it \textit{controlflow-based} serverless computing, including centralized and decentralized controlflow. The differece between centralized system and decentralized system is the scheduler. Centralized system only deploys a global scheduler while decentralized system has a distributed scheduler. Irrespective of the underlying architecture, be it centralized or decentralized, controlflow systems govern function invocation based on function states. This indicates that a particular function is unable to initiate execution until it confirms the successful completion of all its precursor functions.

Existing controlflow-based serverless computing systems suffer from suboptimal performance. In a controlflow-based serverless workflow system, the execution of a function is contingent upon the completion of all its precursor functions. This dependency ensures a sequential progression of tasks, maintaining the integrity of the workflow and its intended outcomes. This invocation pattern can result in increased end-to-end latency, particularly for functions with numerous precursor functions. As each function must wait for its predecessors to complete before initiating execution, the overall workflow execution time may be adversely impacted.

The main insight of this work is that functions can be invoked based on data dependencies rather than function state. This new approach, called the dataflow-based invocation pattern, reduces unnecessary waiting times by allowing functions to be invoked even if their precursor functions are still running. As a result, lower end-to-end latency can be achieved.
It should be noted that the transition from controlflow-based invocation to dataflow-based invocation for serverless workflow execution represents a non-trivial paradigm shift. In order to invoke functions utilizing the dataflow-based invocation pattern with efficiency and minimize superfluous waiting time, we propose the design and implementation of a distributed scheduler named \textit{DScheduler}. Our proposed DScheduler consists of two key components: a global scheduler (GS) deployed on the master node and a dataflow-based local scheduler (DLS) deployed on each worker node. 
The global scheduler in our DScheduler system is responsible for partitioning the workflow into sub-workflows, assigning these sub-workflows to worker nodes. The DLS is designed to invoke functions based on data dependencies between functions, rather than the state of individual functions. In this invocation pattern, a function can be executed even if its precursor functions are still running. Hence, this invocation pattern creates an execution time overlap between function and its precursor functions. The execution order between a function and its precursor functions is not deterministic, and is therefore referred to as out-of-order execution. However, this approach can lead to errors if a function requests data that has not yet been generated by its precursor function. To facilitate out-of-order function correct execution, we have devised a novel \textit{auto blocking/waking up} mechanism incorporated within the DStore system.

In our distributed environment, the serverless workflow is partitioned into multiple sub-workflows and distributed across several machines. Consequently, there are two types of data movement to consider between functions: intra-node data movement for functions located on the same node, and inter-node data movement for functions located on different nodes.

To achieve efficient data movement between functions, we introduce DStore, a distributed in-memory key-value store that provides a user-friendly API and supports updates of arbitrary data sizes. The DStore comprises two key components: (1) a data directory service, and (2) a local store on each worker node. We separate the data and metadata within the DStore. Specifically, the data directory service is architected to store metadata exclusively, incorporating an automatic blocking mechanism that accommodates out-of-order function execution. The local store is specifically designed to store data. By invoking the DStore API, the container responsible for executing a function can ascertain whether the required data is situated on the same node or a remote one. The local store adeptly exploits data locality, substantially reducing intra-node data movement overhead. Additionally, the DStore integrates an efficient fine-grained data retrieval mechanism, which facilitates functions in seamlessly obtaining data from the DStore. To fully utilize network bandwidth, we have proposed a decentralized, receiver-driven coordination mechanism within the DStore.

Building upon the DScheduler and DStore, we propose \textit{DFlow}, an efficient, dataflow-based serverless workflow system that supports complex workflows involving constructs such as "foreach" and others. DFlow utilizes the dataflow-based invocation pattern in conjunction with the DScheduler to invoke functions. This pattern relies on the presence of data dependencies between functions, rather than the state of the functions. This feature enables DFlow to execute a function, regardless of whether its immediate predecessor function remains incomplete or has yet to be executed. As a result, DFlow utilizes the DScheduler to prewarm both the function's container and its precursor functions' containers concurrently. This approach effectively reduces the cold-start latency. DFlow also leverages the DStore to ensure
accurate execution of functions and to promote efficient data exchange among them. This cultivates effective communication and
expeditious data transfer between interdependent functions.

In summary, we make the following contributions.

\textbf{(1)} In this paper, we analyze the limitations of controlflow-based serverless workflow systems, and propose a novel distributed dataflow-based scheduler, called DScheduler. Unlike traditional schedulers that rely on function state, DScheduler leverages dataflow-based invocation pattern to invoke and execute functions.
    
\textbf{(2)} We present the design and implementation of a hierarchical pipelining distributed key-value store, referred to as DStore. DStore enables out-of-order function execution, enhances data locality, and facilitates efficient data transfer among functions.

\textbf{(3)} To showcase the benefits of our proposed dataflow-based approach, we integrate DScheduler and DStore into a serverless workflow system, named DFlow. Through a comprehensive evaluation on representative real-world applications, we demonstrate that DFlow can achieve high throughput and low latency while improving network bandwidth utilization. DFlow can also reduce the cold startup latency.

DFlow demonstrates remarkable performance gains, achieving an average improvement of 60\% over CFlow, 40\% over FaaSFlow, 25\% over FaasFlowRedis, and 40\% over KNIX in terms of 99th-percentile latency. Additionally, it attains higher throughput compared to CFlow, FaaSFlow, FaaSFlowRedis, and KNIX. DFlow also enhances network bandwidth utilization by 2.0x-4.0x over CFlow and 1.5x-4.0x over FaaSFlow ~\cite{FaaSFlow}, FaaSFlowRedis ~\cite{FaaSFlow}, and KNIX ~\cite{SAND-ATC18}, respectively. Furthermore, DFlow significantly reduces cold startup latency, exhibiting an average improvement of 5.6x over CFlow and 1.1x over FaaSFlow.

 \begin{figure}[htp]
     \centering
     \includegraphics[width=\linewidth]{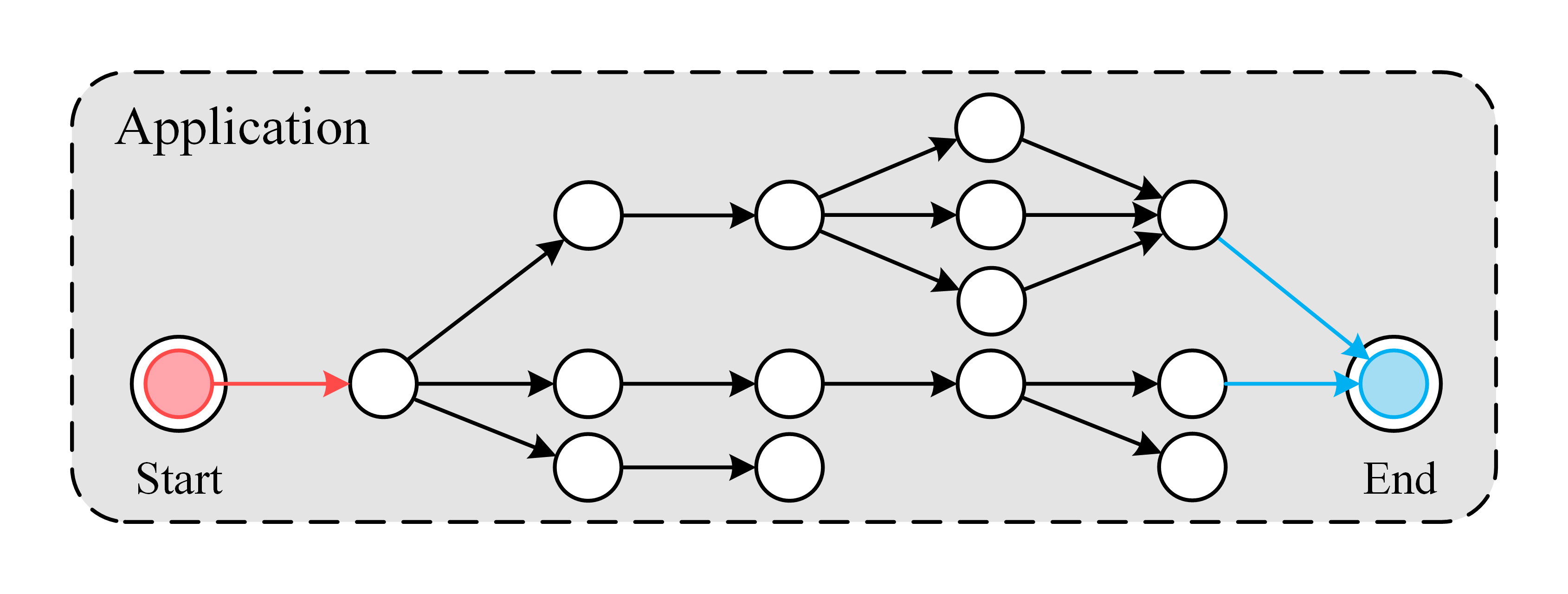}
     \caption{Example of serverless workflow that contains multiple functions and its form is not a simple chain.}
     \label{fig:workflow}
 \end{figure}

\begin{figure*}[htb]
\centering
\label{fig:clow-ce-arch}
\subfigure[\textbf{CFlow}]{
\includegraphics[width=0.45\textwidth]{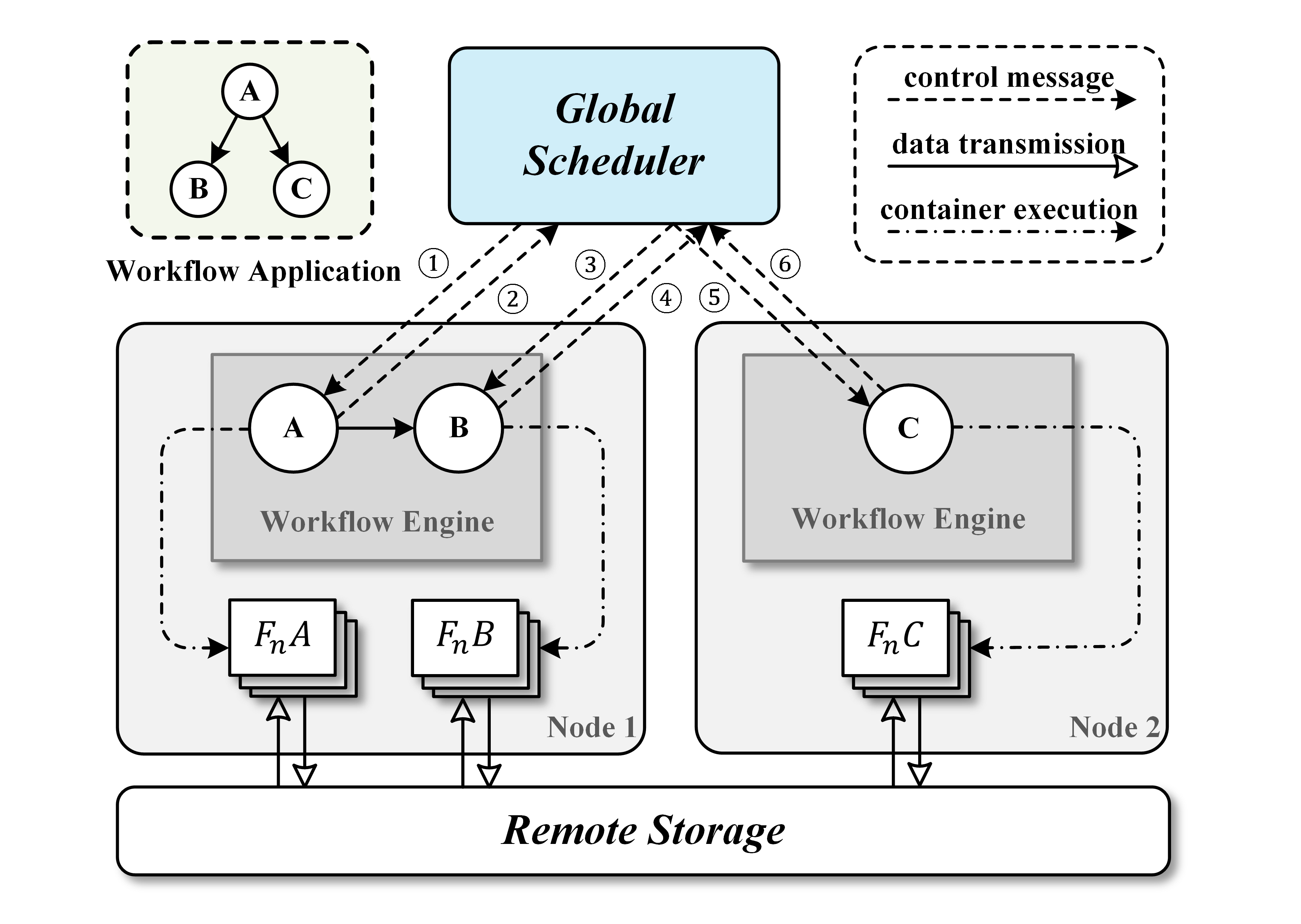}
}
\label{fig:cflow-arch}
\subfigure[\textbf{FaaSFlow}]{
\includegraphics[width=0.45\textwidth]{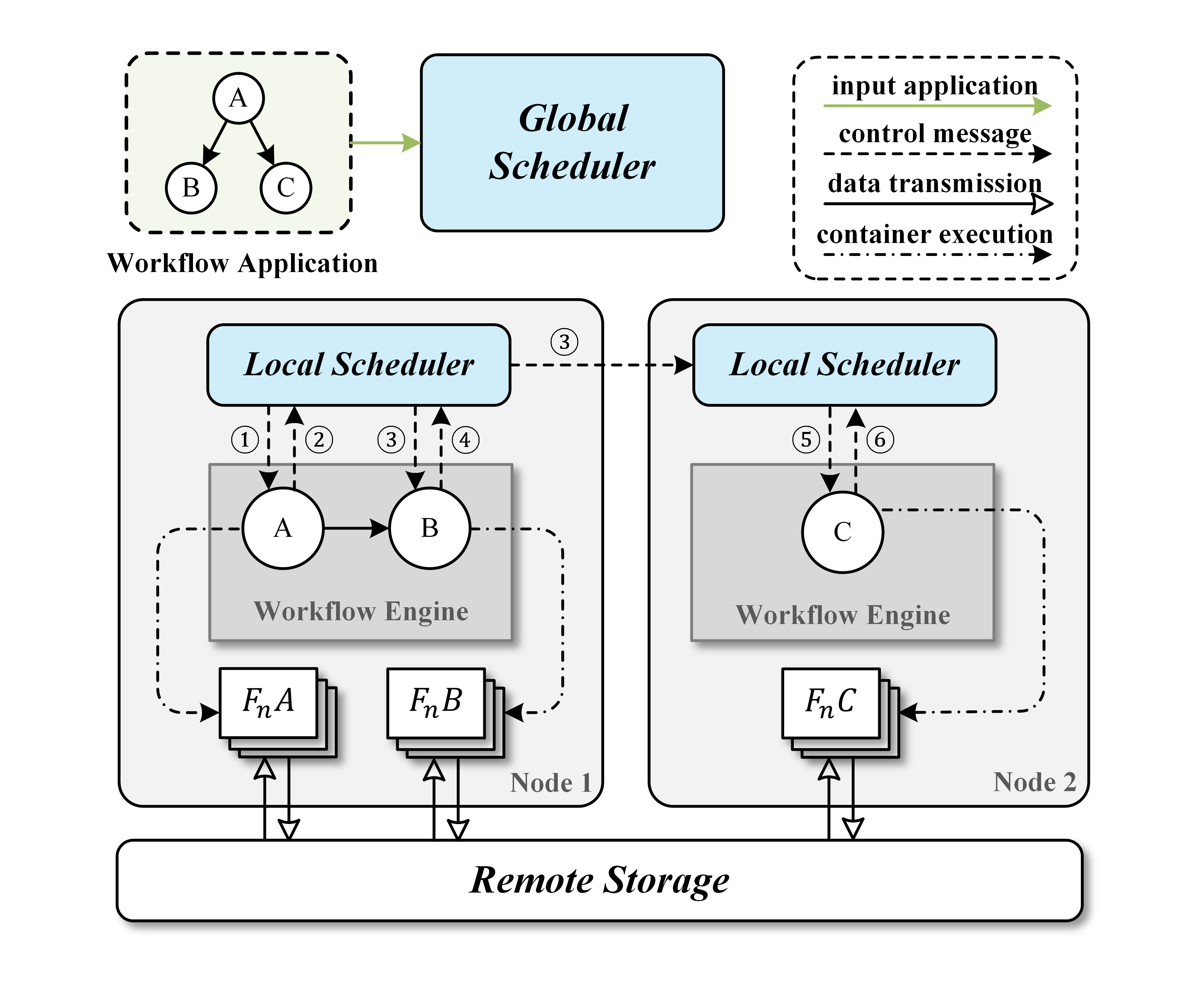}
}
\label{fig:clow-de-arch}
\caption{The subfigure (a) is  the architecture of Centralized-flow(\textbf{CFlow}).The subfigure (b) is the  architecture of de-centralized-flow(\textbf{FaaSFlow}). Both of then are controlflow-based serverless system.}
\label{fig:controlflow}
\end{figure*}
\section{Background and Motivation}
\label{sec:background}
\subsection{Serverless Workflow}
An application can be represented as a directed acyclic graph (DAG). Within the context of serverless computing, the DAG is commonly referred to as a serverless workflow, with each node representing a function and each edge representing a data dependency between a source and a destination function. The source function operates as the precursor to the destination function, and conversely, the destination function serves as the successor of the source function.

 The simplest form of serverless workflow is the sequential workflow, which can be viewed as a function chain. The Sequoia system~\cite{Sequoia} focuses on the function chain and implements a chain scheduler to ensure Quality of Service (QoS). Recent work~\cite{FaaSFlow, Boki-SOSP21, cloudburst} not only supports the sequential workflow but also supports more complex workflows. For example, Figure~\ref{fig:workflow} shows a non-trivial workflow containing complex functions, which is not in the form of a simple function chain.
\subsection{Controlflow-based Serverless Workflow System }
\subsubsection{\textbf{Invocation Pattern}} \

Previous controlflow-based serverless systems, such as those presented in~\cite{FaaSFlow,Nightcore-ASPLOS21,SONIC,Faastlane,SAND-ATC18}, can be categorized into two types: centralized and decentralized systems. The architecture of the two types of systems is illustrated in Figure \ref{fig:controlflow}. Specifically, functions A and B are located on worker node 1, while function C is on worker node 2. Functions B and C depend on the output of function A, which serves as a required input for both. The key difference between the centralized system (CFlow) and the decentralized system like FaaSFlow~\cite{FaaSFlow} is the scheduler. CFlow deploys a global scheduler, while FaaSFlow deploys a two-layer scheduler.

In CFlow (Figure \ref{fig:controlflow}a), the global scheduler on the master node partitions the DAG into sub-DAGs and assigns them to different worker nodes. Additionally, the global scheduler is responsible for managing the function execution state by sending invocation messages to worker nodes to invoke functions. As illustrated in CFlow (Figure \ref{fig:controlflow}a), the Global Scheduler initially sends an invocation message to Node 1 to commence the execution of function A. Upon confirming the completion of function A, the Global Scheduler proceeds to send invocation messages for functions B and C, respectively. 


In FaaSFlow (Figure \ref{fig:controlflow}b), a global scheduler on the master node and a local scheduler on each worker node are deployed. The global scheduler partitions the DAG into sub-DAGs and assigns them to different worker nodes. The local scheduler on each worker node is responsible for managing the function execution state by sending control messages to invoke functions. In Figure \ref{fig:controlflow}b, the local scheduler on worker node 1 (LS1) initiates the execution of function A. After completion of function A, LS1 starts the execution of function B while simultaneously sending the invocation message to LS2. Upon receiving the invocation message, LS2 starts the execution of function C.


 \begin{figure}[htbp]
     \centering
     \includegraphics[width=\linewidth]{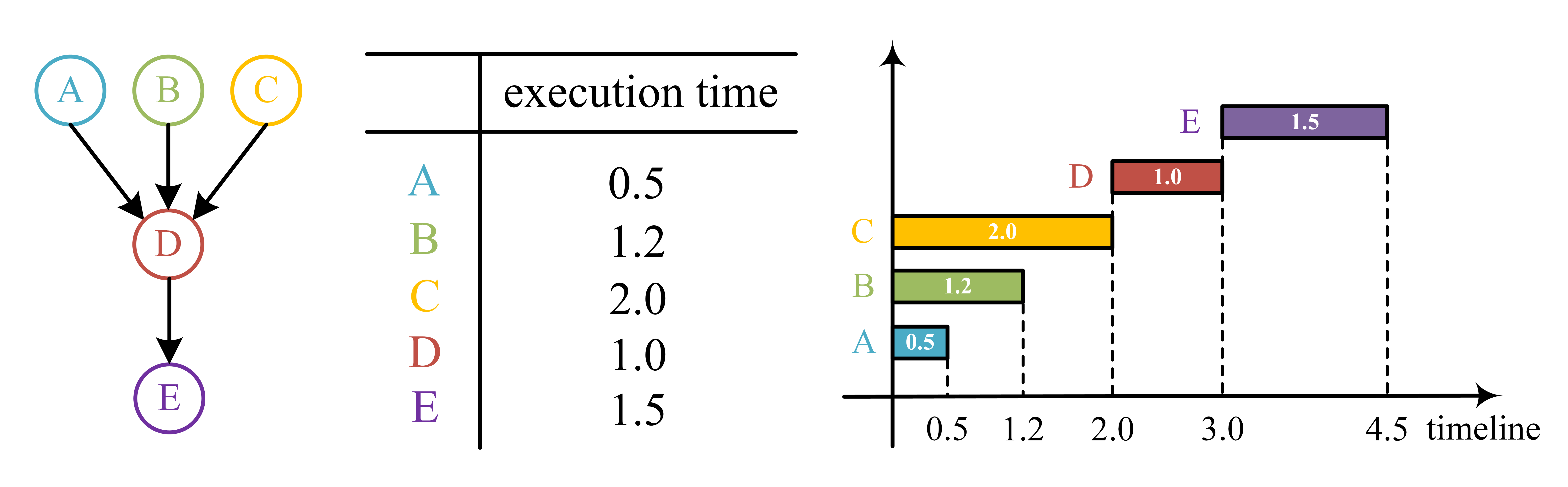}
     \caption{ The function execution timelime in controlflow-based serverless workflow system. Function D needs the output of function A, B, C as input. Function E needs the output of function D as input. Function A, B, and function C begin their execution at 0s. Function D starts at 2.0s, while E commences execution at 3.0s.}
      \label{fig:control-flow-invocation}
 \end{figure}

The defining characteristic of controlflow-based serverless systems is the function invocation pattern, which relies on the state of predecessor functions to trigger execution. In particular, if a function has precursor functions, it can only commence execution after ensuring that all its precursors have finished executing. Figure \ref{fig:control-flow-invocation} presents an illustrative execution timeline of this invocation pattern, featuring a serverless workflow in which functions A, B and C are precursors of function D, and function D is a precursor of function E. In a controlflow-based system, it is essential to note that function D commences execution only upon the completion of functions A, B, and C, while function E initiates its execution subsequent to the termination of function D.

\subsubsection{\textbf{Data-Shipping Pattern}} \

\begin{figure}[htbp]
     \centering
     \includegraphics[width=\linewidth]{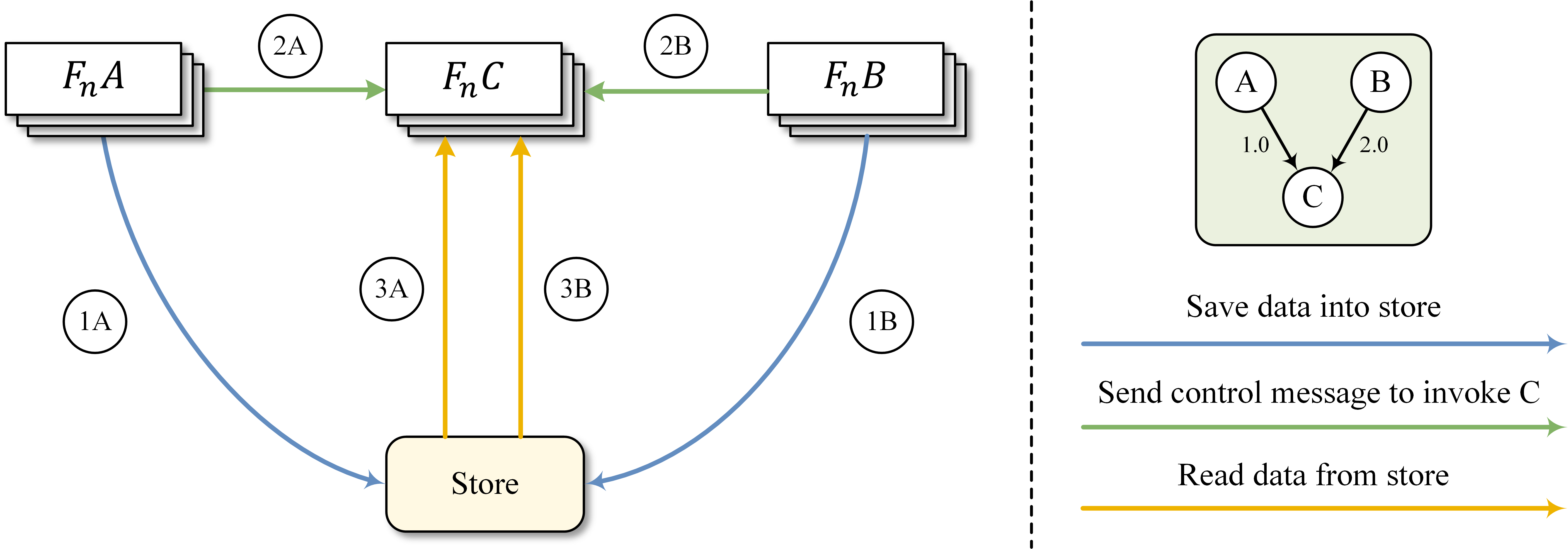}
     \caption{ The data-shipping pattern in controlflow-based serverless workflow system. }
      \label{fig:control-flow-data-shipping}
 \end{figure}

Serverless computing is an architecture based on data shipping~\cite{ServerlessStep}, where the output of a function is shipped to its precursor functions in a serverless workflow. Platforms such as OpenWhisk~\cite{OpenWhisk, AWS-lambda} use a storage system to exchange data. Figure~\ref{fig:control-flow-data-shipping} shows the data-shipping pattern of a controlflow-based serverless system, where function C necessitates the utilization of the outputs generated by both function A and function B as its input parameters. Initially, upon completion of its execution at the 1s mark, Function A stores its output in store (Step 1A). Subsequently, it transmits a control message in an attempt to initiate function C(Step 2A). However, function C is unable to commence execution as function B remains active. Once function B concludes its operation at the 2s mark, it proceeds to save the generated data into store(Step 1B), followed by sending a control message to invoke function C(Step 2B). Consequently, function C is now able to initiate execution and retrieve the necessary data from store.



 Controlflow-based serverless systems suffer from suboptimal performance.
 This is because the system invokes functions based on their state with controlflow invocation pattern. However, this pattern will cause long end-to-end latency~\cite{ServerlessComputingSurvey} because of the waiting time, especially for functions with many precursor precursor functions (e.g., Function D in Figure~\ref{fig:control-flow-invocation}). Considering the limitations of the controlflow-based invocation pattern, it is essential to explore a novel invocation pattern where a function can start executing even if some of its precursor functions are still running. We call this pattern the dataflow-based invocation pattern, and the system is called the dataflow-based serverless workflow system.

 \subsection{\textbf{Challenges in DataFlow-based Invocation }}
It can be highly challenging to realize a dataflow-based serverless workflow system for several reasons. \textbf{First}, it is essential to develop a scheduler capable of handling the dataflow-based invocation pattern, ensuring the ability to invoke a function even if its precursor functions are still executing. \textbf{Second},  given the non-deterministic execution order of functions and their precursors in a dataflow-based invocation pattern, it is crucial to guarantee that a function receives the correct data as input, regardless of whether some of its precursor functions are still running. \textbf{Third},  in a distributed environment where the workflow DAG is partitioned into several sub-DAGs distributed across multiple machines, it is vital to leverage data locality for efficient intra-node data exchange and optimize network bandwidth usage for effective inter-node data exchange.



 \begin{figure}[ht]
     \centering
     \includegraphics[width=\linewidth]{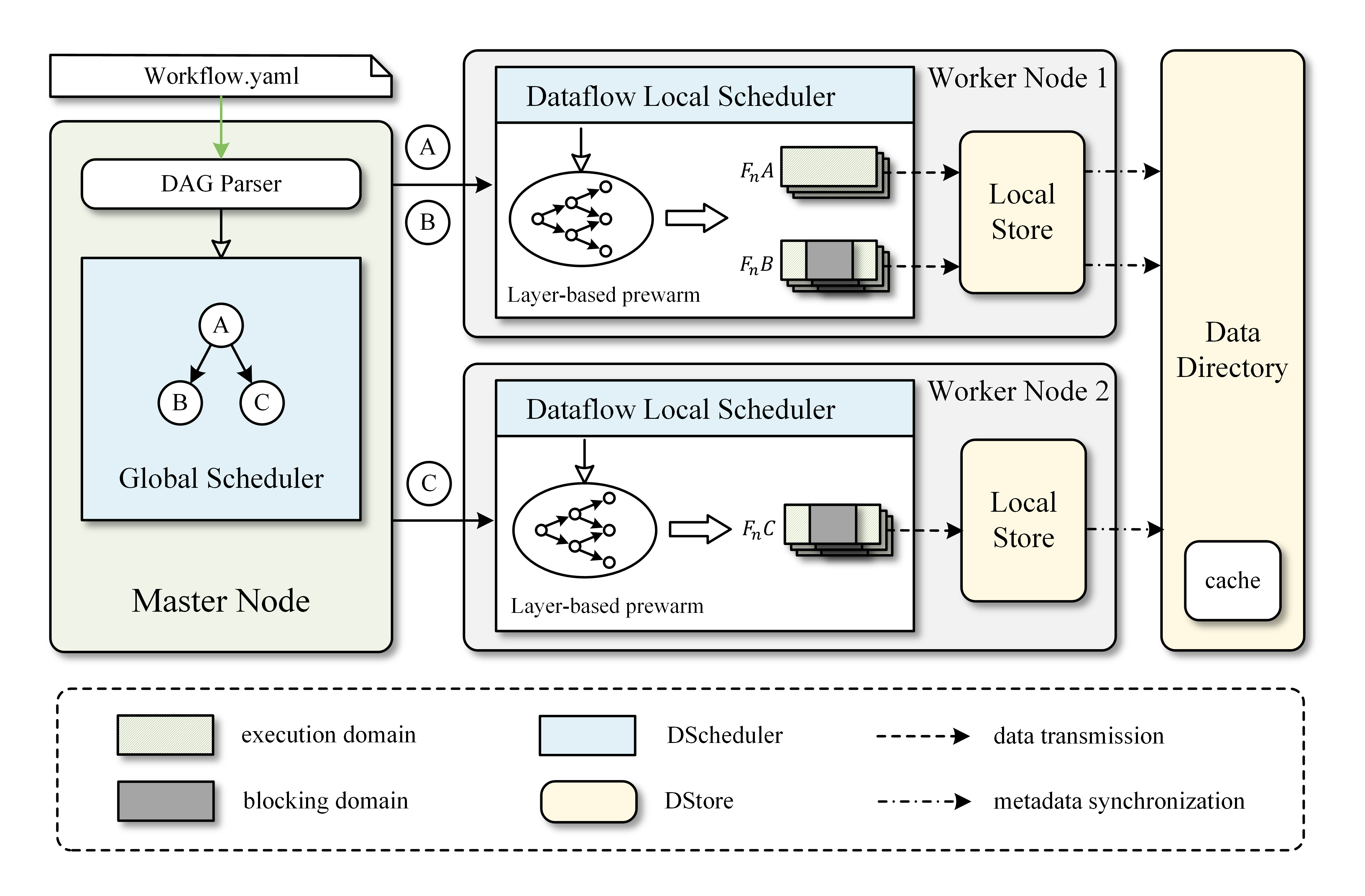}
     \caption{ The system overview of DFlow. }
      \label{fig:DFlow-overview}
 \end{figure}

\section{System Design of DFlow}

\subsection{Overview}


Figure \ref{fig:DFlow-overview} shows the overview of DFlow, a dataflow-based serverless system for serverless workflow. DFlow is composed of three key components: (1) a DAG parser responsible for parsing user configuration files, (2) a dataflow-based DScheduler that invokes functions based on the data dependency between them, and (3) a DStore that enables the exchange of data between functions. DScheduler is a distributed scheduler designed to address \textbf{Challenge 1}. DScheduler comprises a two-tiered architecture, featuring a Global Scheduler (GS) situated on the master node and a Dataflow-based Local Scheduler (DLS) deployed on each worker node.  The DStore is a distributed kv store to tackle \textbf{Challenge 2} and \textbf{Challenge 3}. DStore is structured with a data directory service hosted on the master node, responsible for managing metadata, and a Local Store (Store) present on each worker node, dedicated to data storage. 

In Figure \ref{fig:DFlow-overview}, firstly user uploads the \textit{workflow.YAML} to the master node. The YAML file defines serverless workflow information, includes the functions, the functions' input name, the functions' output name and so on. We call workflow information as workflow metadata. DFlow uses the DAG Parser to parses this configure file into a DAG. The node of DAG represents a function and the edge represents data dependency between source function and destination function. The input of destination function is the output of source function. Then DFlow leaverages the GS to partitions the DAG into sub-DAG and assigns the sub-DAG to different worker node and sends the workflow metadata, such as the entry points of workflow, the function's successor functions to every local scheduler. In Figure \ref{fig:DFlow-overview}, only the function A is the entry point. Function  A and function B are on the worker node1 and function C is on the worker node2. 

Upon receiving a workflow invocation message from event triggers, DLS1 collaborates with DLS2 to simultaneously invoke functions A, B, and C. All these functions are started to prewarm the container and are executed in the container. Function A serves as the entry point of the workflow. Upon completion of its execution, it stores the output in the Local Store (Store1) and writes the corresponding metadata into the data directory service. To ensure that B and C can access the correct data, the auto waking/blocking mechanism temporarily blocks the execution of B and C until the required metadata becomes available in the data directory service. Upon waking up, both  B and C retrieve the necessary data with the assistance of DStore and subsequently continue their execution. Upon completion of the execution of all functions within the workflow, their resultant output data and associated metadata are persisted in DStore.

Through the above steps, the example workflow is executed by a dataflow-based architecture.

\begin{algorithm}[t]
 \caption{dataflow-based local scheduler(DLS) logic } 
 \KwIn{entry\_point\_funs: entry points functions of this Node } 
 
    \For{f in entry\_point\_funcs}{
        run(f)  
        
        next\_funcs =f.next  
        
        \For{next\_f in next\_funcs}{
            run(next\_f) 
        }
    } 
    
    \tcp*[l]{when one of function is done.}
    \For{f in entry\_point\_funcs } {
        next\_funcs = f.next  \\ 
        \For{next\_f in next\_funcs}{ 
            next\_next\_funcs = next\_f.next  \\ 
                \For{next\_next\_f in next\_next\_funcs}{
                    run(next\_next\_f)  
            }
        }
    }
        
 \end{algorithm}
 \subsection{DScheduler}
The DScheduler is a distributed scheduler that contains a global scheduler on master node and each work node has a dataflow-based local scheduler(DLS). The global scheduler partitions a workflow into sub-workflows and assigns them to worker nodes. Furthermore, it propagates essential workflow metadata, such as entry points and function input IDs, to each local scheduler, enabling them to comprehend the data interdependencies between functions for subsequent dataflow-based invocations.

In DFlow, a function can be invoked to execute even if its precursor functions are still running. This invocation pattern is referred to as dataflow invocation. Additionally, the execution of functions in DFlow occurs out-of-order. Algorithm 1 illustrates the dataflow-based schedule for each local scheduler.



 When the DLS receiving a workflow invocation by event triggers, each DLS knows the metadata of this workflow, such as the entry points, the entry points' successor functions and others. Initially, the DLS iterates through the list of entry points and begins executing them concurrently(lines 1-2). For each entry point, the DLS retrieves its successor functions, iterates through them, and invokes them (lines 3-6) to execute simultaneously, even if the entry point function is still running. When one of the entry points finishes, the DLS retrieves its successor's successor function and invokes it(lines 8-15) to execute. For any function that has completed execution, the DLS repeats lines 8-15 to execute the remaining functions until all the functions of the workflow have been finished. 
  \begin{figure}[h]
     \centering
     \includegraphics[width=\linewidth]{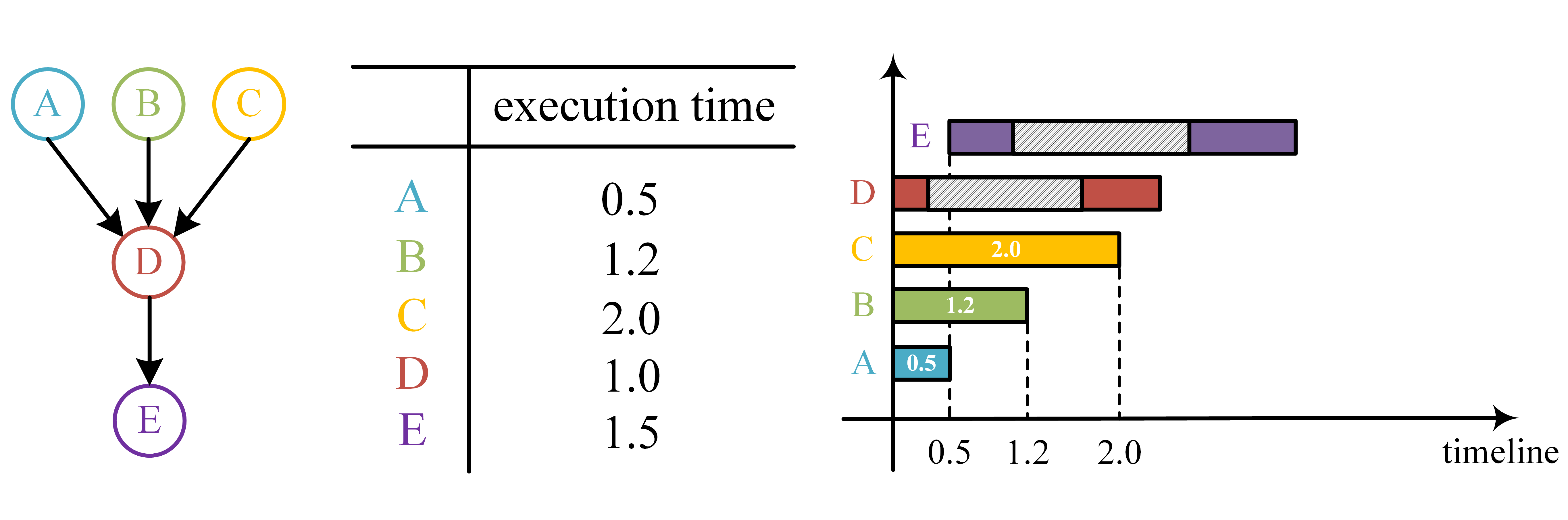}
     \caption{An example of the dataflow-based invocation pattern used in DFlow, consider a workflow with precursor functions A, B, and C leading to function D, which in turn is a precursor to function E. DLS invokes functions A, B, C, and D simultaneously at 0s. Once function A completes at 0.5s, it triggers the immediate invocation of function E.}
      \label{fig:dataflow-invocation}
 \end{figure}
For instance, as depicted in Figure~\ref{fig:dataflow-invocation}, DLS simultaneously initiates the execution of functions A, B, C and D at time 0s. Upon completion of function A, it promptly proceeds to invoke function E at time 0.5s.

The DLS employs a dataflow-based invocation pattern to invoke functions, which can introduce an execution time overlap between a function and its precursor functions (e.g., Function D and functions A, B, Cin Figure~\ref{fig:dataflow-invocation}), leading to a reduction in cold-start latency. The DLS can be designed to invoke all functions of a workflow immediately upon receiving a workflow invocation from event triggers, which can achieve the fastest cold-start latency. However, this approach can also result in inefficient resource utilization. Functions that depend on precursor functions must wait for the required data to be generated, even if their precursors are still executing. Functions with numerous precursor functions, including those with precursor functions of their precursors, may experience prolonged wait times, leading to suboptimal resource utilization. DScheduler balances the trade-off between minimizing cold-start latency and optimizing resource utilization. In DScheduler, DLS initiates the invocation of entry points and their successor functions(line 1-7 Algorithm1). Once one of the above functions is completed, it only activates the finished function of successor function of successor functions(line 10-14).

\begin{table}[h]
\caption{The Core API of DStore}
\centering
\resizebox{0.75\textwidth}{!}{
\begin{tabular}{c|c}
\hline
\thead{Name} & \thead{Description} \\ \hline
 \thead[l]{Buffer buffer <- Get(string key)} & \thead[l]{Return a data buffer with the key. \\ This can be blocking.} \\ \hline
\thead[l]{Put(string key, Buffer buffer)} & \thead[l]{Create a data with the given key and buffer}\\ \hline
\end{tabular}}

\end{table}

\subsection{DStore}

DStore is a distributed store for dataflow-based serverless system. Table 1 is its core API. In the DFlow system, the function retrieves data from the DStore by invoking the \textit{Get} method. If modifications are made to the data, the updated version must be stored in the DStore by calling the \textit{Put} method with a new, unique identifier. This ensures that each distinct data item within the DStore is associated with its own unique ID. At a high level, DStore uses these techniques:1)data directory service that save metadata; 2)auto blocking/waking up mechanism to enable correct dataflow-based function execution; 3)fine grained data exchange to reduce the tail latency; 4)decentralized Receiver-driven mechanism to fully utilize the network bandwidth.

\subsubsection{ \textbf{Data Directory Service}}
\



The data directory service maintains the metadata for each data: 1)the ID of the data; 2)the size of the data; 3) the location information; 4)the data location's access frequency. The other node can write the metadata to the data directory service when a process creates a data by calling Put on DStore or when the data is copied from other node. Since the metadata size is minimal, the time taken to write it into the data directory service is approximately 150us. As a result, the data directory service can maintain multiple metadata entries for a single piece of data once it has been transferred to other nodes.

The location information is a list of IP addresses that indicate which node contains the data. In serverless workflows, a potential scenario may arise where a function, denoted as  A on Node 1, has numerous successor functions located on different nodes . Because all of the successor functions need function A's output as input, the node 1 has to send the data to many node and node1 may become a performance bottleneck. 
To mitidate this situation, the data directory service maintains multiple metadata entries for the data once it has been transferred to various nodes. The data directory service keeps track of the frequency of metadata accesses for each node. When a client queries this metadata, the data directory service responds by providing the server IP with the lowest access frequency. 

\subsubsection{ \textbf{Auto Waking-up/Block Mechanism}}  \


The DScheduler is designed to invoke a function even while its precursor functions are still executing, allowing for out-of-order function execution between a function and its preceding counterparts. To ensure that a function obtains the correct data from its precursor functions, the data directory service integrates an automatic wake-up/blocking mechanism. If the function cannot find the required data in its local store, it will query the data directory service for the corresponding metadata. In cases where the metadata is unavailable in the data directory service, the function is blocked by the automatic wake-up/blocking mechanism. Once the precursor function writes the necessary metadata to the data directory service, the mechanism wakes up the blocked function, enabling smooth execution.

\subsubsection{\textbf{Fine Grained Data Retrieve}} \


 \begin{figure}[ht]
     \centering
     \includegraphics[width=\linewidth]{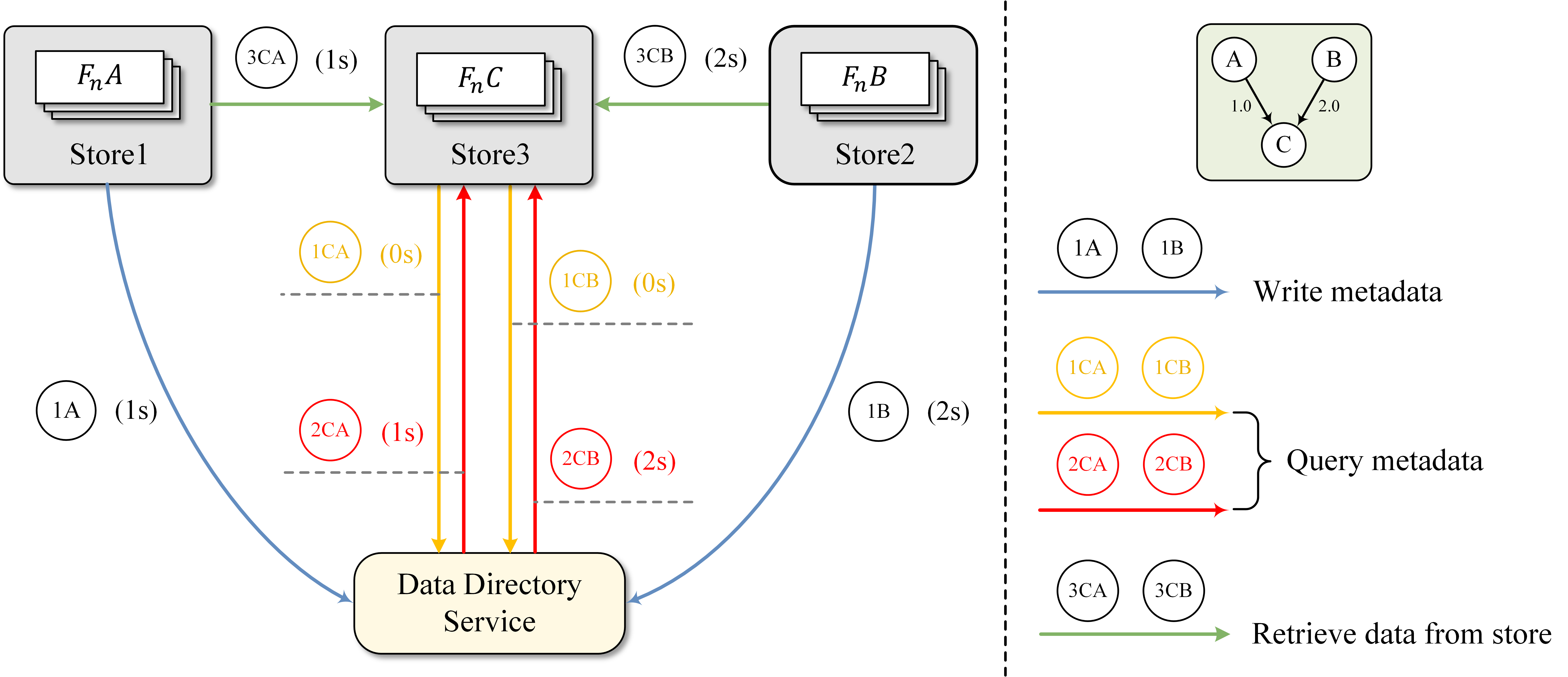}
     \caption{ The data shipping pattern in DFlow. The execution time of A is 1s and execution time of B is 2s.  function C requires the outputs from
functions B and A as input. }
      \label{fig:dataflow-shipping}
 \end{figure}

As illustrated in Figure~\ref{fig:dataflow-invocation}, there exists an execution time overlap between function D and function A, B, and C. Leveraging the execution time overlap and aiming to minimize data movement latency, DStore implements a fine-grained data retrieval mechanism specifically designed to efficiently reduce tail latency. We use an example to show this  fine-grained data retrieval mechanism in Figure~\ref{fig:dataflow-shipping}.

As depicted in Figure~\ref{fig:dataflow-shipping}, function C requires the outputs from functions A  and B as input. At time 0s, functions A, B , and C are invoked simultaneously. Following this, C  initiates two threads (A, B) to query the data directory service for metadata. However, these two threads become blocked since the required metadata is not yet available in the data directory service (Step 1CA, 1CB figure~\ref{fig:dataflow-invocation}). Upon completion of A's execution and subsequent metadata writing to the data directory service at 1s(Step 1A, Figure~\ref{fig:dataflow-invocation}), thread A is awakened, while thread B remains blocked(Step 2CA Figure~\ref{fig:dataflow-invocation}).Subsequently, A proceeds to retrieve the pertinent data from the Store1(Step 3CA, Figure~\ref{fig:dataflow-invocation}). Subsequently, B awakens once B finishes its execution at 2s(Step 1B, Step 2CB Figure~\ref{fig:dataflow-invocation}), following a similar process to that of A(Step 3CB Figure~\ref{fig:dataflow-invocation}).  By employing the aforementioned fine-grained data retrieval mechanism, the execution time overlap can be leveraged to optimize data exchange overhead. As a result, this approach contributes to a reduction in tail latency.


\subsubsection{\textbf{Receiver-Driven for Inter-Node Function}} \


DStore implements a receiver-driven coordination mechanism for inter-node function data exchange. In DStore, inter-node data transfer over the network happens only in one scenarios: the function finds the data it wants is on  the remote node.

In serverless environment, it's normal that the two functions that exist data-dependency are on the different modes. When a function requires the output of its precursor function as input, it first checks whether the data is available in its local store. If the data is not present, the function queries the data directory service to obtain the location information of the required data. By utilizing the decentralized receiver-driven coordinator mechanism, functions can directly fetch data from remote nodes, enabling them to maximize the use of available network bandwidth and optimize data transfer efficiency throughout the network.

\subsubsection{\textbf{Fault Tolerance}} \


Currently, DFlow primarily concentrates on the impact of the invocation pattern for workflow execution, without addressing fault tolerance aspects. If one node fails, DFlow utilizes global scheduler to re-partitions the workflow and re-execute the workflow from the beginning. And the DStore will delete the previous data and metadata.

\subsection{\textbf{Putting Together}}

\begin{figure}[t]
	\centering    
	\subfigure[The code of function, function B and function C.] 
	{
   	   \includegraphics[scale=0.5]{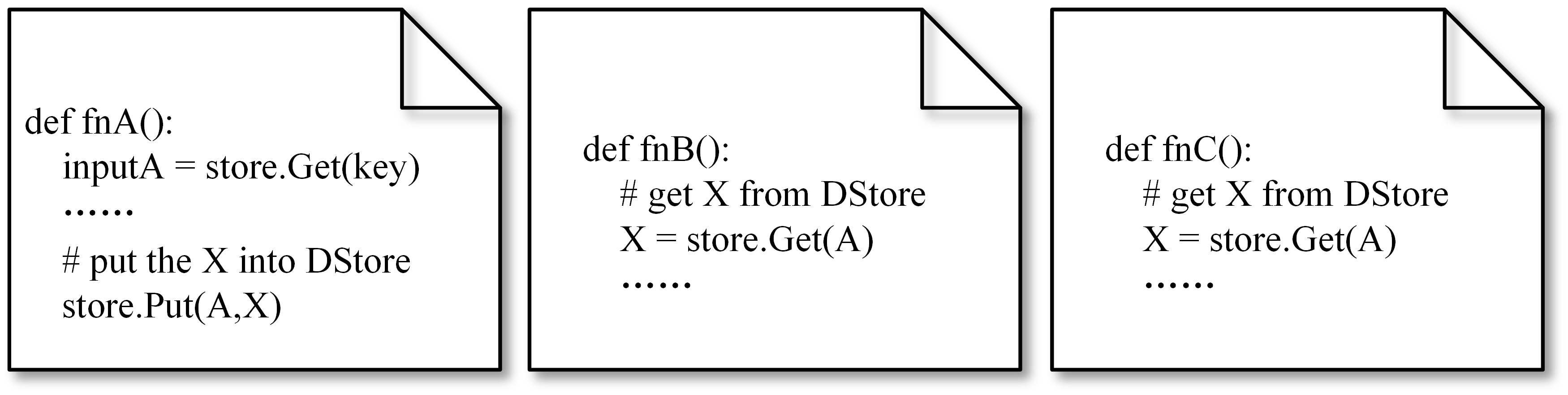}

	}
	 \vspace{-2mm}
	\subfigure[Function execution with the support of DStore] 
	{       
 		\includegraphics[scale=0.4]{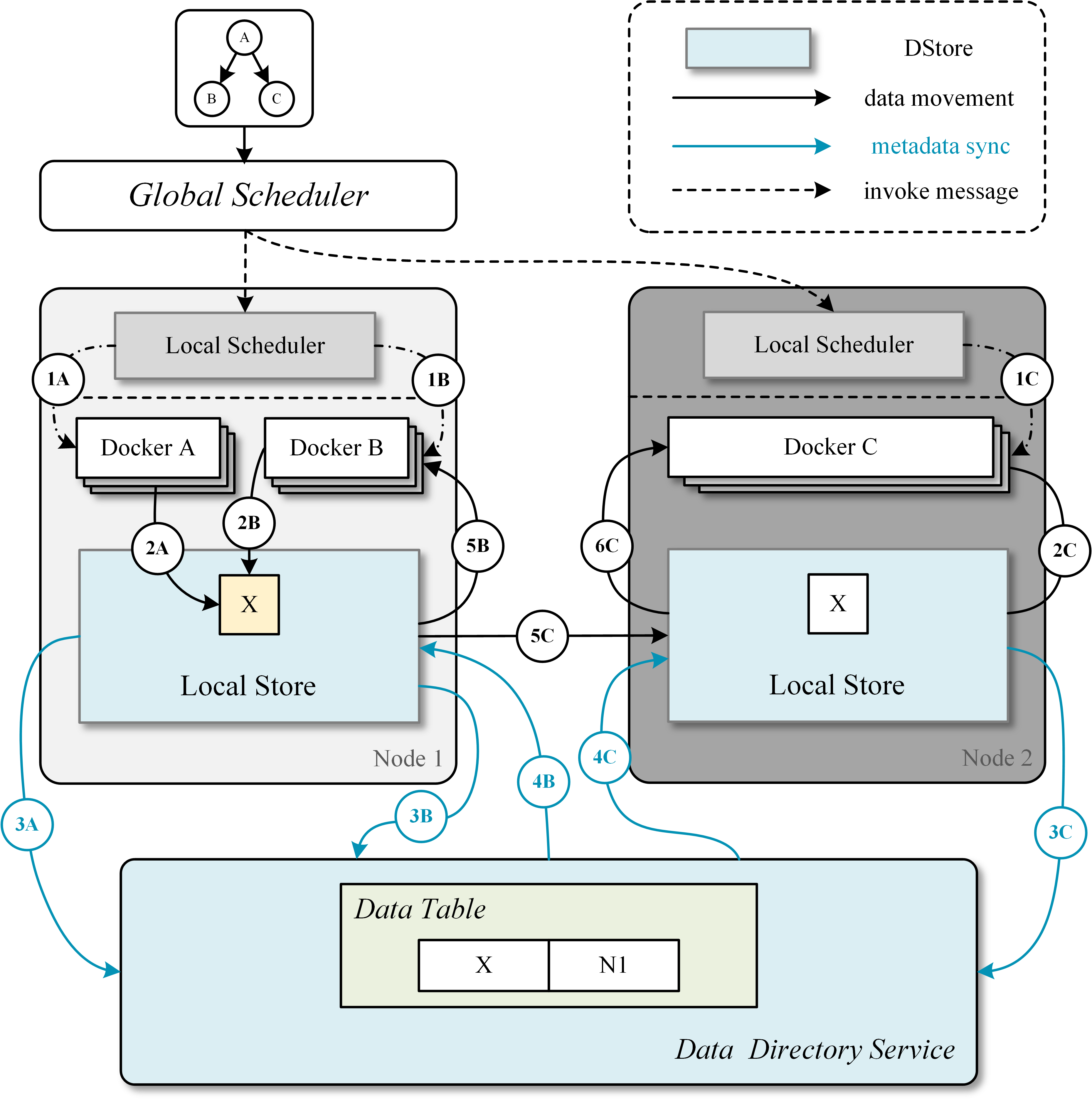}
}
\caption{Example of simple serverless workflow.The system contains a pool of docker per physical node and a global scheduler to schedule  function to different node. Every DStore consists of one local store per work node and  global distributed  data directory service, which is distributed across physical nodes. Every node uses a docker to execute a function. The subfigure (a) is the function code of this workflow. The subfigure (b) is the procedure this workflow by using DStore for data transfer.}
\label{fig:example-Dstore}
\end{figure}

It is non-trivial to put the above techniques together in a system to support a dataflow-based invocation for serverless workflow. We use an example to illustrate how DFlow effectively coordinates the DScheduler and DStore to enable dataflow-based invocation pattern.

In our example, we use the core API of DStore (see Table 1) to implement a simple serverless workflow comprising three functions, as depicted in Figure ~\ref{fig:example-Dstore} a). Functions B and C rely on the output of function A, which we refer to as data X.

At the beginning, the global scheduler partitions the serverless workflow by assigning function A and function B to Node 1, while placing function C on Node 2. Upon receiving a workflow invocation triggered by an event, the local scheduler of Node 1 (DLS1) simultaneously utilizes Docker A to execute function A  and Docker B to execute function B. At the same time, DLS1 sends a message to the local scheduler of Node 2 (DLS2) to invoke function C (refer to steps 1A, 1B, and 1C in Figure \ref{fig:example-Dstore}). Since Docker B and Docker C require data X as input, they call the Get(X) operation on their local store, respectively (see steps 2B and 2C). If the local stores of Node 1 (Store1) and Node 2 (Store2) do not have data X, they will query the data directory service for its metadata, as depicted in steps 3B and 3C of Figure \ref{fig:example-Dstore}. If the data directory service lacks the metadata for data X at this moment, the automatic waking-up/blocking mechanism will block processes B and C until the required metadata becomes available.



While docker A executes function A, it calls Put(X) on DStore when function A finishes. Then, docker A copies data X to Store1 (step 2A in Figure \ref{fig:example-Dstore}). This frees docker A for next reuse , reducing cold start latency for subsequent workflow invocations. Concurrently, Store1 writes the metadata of X to the data directory service asynchronously (step 3A in Figure \ref{fig:example-Dstore}).




Once the data directory service has the metadata for data X, it will automatically wake up processes B and C and provide them with the required metadata (see steps 4B and 4C in Figure \ref{fig:example-Dstore}). 

Process B knows that data X is located in its local store, process B can immediately copy X from Store1 in step 5B. Process C, on the other hand, knows that data X is located on remote node1. In step 5C, Store2 copies data X from Store1 over the network. Finally, in step 6C, docker C copies data X from Store2 into its memory and successfully executes function C.

\section{Implementation}


We built DFlow on top of Hoplite{~\cite{Zhuang2021Hoplite} and FaaSFlow{~\cite{FaaSFlow}}. We implement DFlow's key design(Figure ~\ref{fig:DFlow-overview}) with approximately 3k lines of C++ and 2K lines of python code.

The global scheduler in the DScheduler is similar to the global
scheduler found in FaaSFlow. Additionally, the DAG Parser shares similarities with the DAG Parser employed in FaaSFlow.The local scheduler is implemented with C++ code and it uses the wfrest~\cite{wfrest} to receive(send) messages from(to) other nodes. We remove all relative collective communication code from Hoplite to implement DStore. The inter-node pipeline mechanism in Hoplite is not used for any benchmark because the output of a single function is at most tens of MB. This relatively small size eliminates the need to use gRPC Stream for slicing the data into multiple pieces. The per-node local store is implemented using a set of gRPC components~\cite{grpc}, functioning as both a gRPC server and a gRPC client. In its capacity as a gRPC server, it processes requests from Docker containers to read data from the local store via the Get method or write data to the local store using the Put method. As a gRPC client, the component communicates with the data directory service to either write data metadata or query the service for metadata information. The data directory service is implemented as a gRPC server using a set of gRPC components  and is distributed
throughout the cluster. As a gRPC server, it processes requests from the per-node local stores, performing operations such as updating metadata for a specific data item or querying metadata for a particular data item. The auto waking-up/blocking mechanism is implemented by integrating gRPC with C++ multithreading.

\section{Evaluation}
In this section, we evaluate the performance of DFlow in comparison to four controlflow-based serverless workflow systems, namely CFlow, FaaSFlow~\cite{FaaSFlow},FaaSFlowRedis~\cite{FaaSFlow}, and KNIX~\cite{SAND-ATC18}. The benchmarks in Table \ref{tableZ:benchmark} are  from FaaSFlow\cite{FaaSFlow, Pegasus1,Pegasus2}, including 2 real-world application and 4 scientific calculation application. All the calculation applications consist of more than 40 functions each.


    
\begin{itemize}[noitemsep,topsep=0pt]
    \item \textbf{CFlow:} This centralized controlflow-based serverless workflow system utilizes CouchDB~\cite{CouchDB} for exchanging all data. Its architecture is depicted in Figure~\ref{fig:controlflow}(a). 
    \item \textbf{FaaSFlow:} This decentralized controlflow-based workflow system employs a hybrid store for data exchange. It utilizes Redis~\cite{redis} for intra-node data exchange and CouchDB~\cite{CouchDB} for inter-node data exchange. The architecture of FaaSFlow is illustrated in Figure~\ref{fig:controlflow}(b).
    \item \textbf{FaaSFlowRedis:} This decentralized controlflow-based workflow system leverages a hybrid Redis architecture for data exchange. A Redis instance is deployed on the master node for inter-node data exchange, while separate Redis instances are deployed on each worker node for intra-node data exchange. The architecture of FaaSFlowRedis is illustrated in Figure~\ref{fig:controlflow}(b)
    \item \textbf{KNIX:} This controlflow-based system relies on different processes to execute different function within one container. Similar to FaaSFlowRedis, it adopts a hybrid Redis architecture for data exchange. The global scheduler from FaaSFlow~\cite{FaaSFlow} is employed to partition the workflow and determine the storage type for each function. For functions with the DB storage type, remote Redis is used for data exchange, while for functions with the MEM storage type, local Redis is utilized for data exchange.
\end{itemize}

\subsection{Experiment Setup} \label{benchmarkExp}

All of experiments are executing on the Aliyun EC2 instances. We set up a 8-node cluster on aliyun to run  all experiment. All the nodes use ecs.g7.2xlarge instances, each of which has 8vCPUs and 32GB of DRAM  and 100 GB ESSD. The software environment is in the Table \ref{table:envirment}.

\begin{table}[t]

\centering
\begin{tabular}{r|l}
\hline
 & \thead{Configuration} \\ 
 \hline
 \thead{SoftWare} & 
 \thead[l]{Ubuntu18.04 \\ 
 Gevent: 21.1.2, gRPC: v1.31.0, Protobuf: v3.12.2 \\ 
 Database: Apache/CouchDB:3.1.1, Redis:6.2.5 }\\ 
  \hline
 \thead{Container} &  
 \thead[l]{Container runtime: Python-3.7.0, Linux with kernel 4.15.7 \\ 
 Resource limit and Lifetime: 1-core with 256MB, 600s  } \\ 
 \hline
\end{tabular}
\caption{The software environment of experiment}
\label{table:envirment}
\end{table}

\begin{table}[h]
\caption{Workflow Benchmark}
\centering
\resizebox{0.75\textwidth}{!}{
\begin{tabular}{c|c}
\hline
\thead{Category} & \thead{Benchmar Name}  \\ \hline
 \thead[l]{Real World Application} & \thead[l]{Word Count(WC) , File Processing(FP).} \\ \hline
\thead[l]{Scientific Calculation Application} & \thead[l]{Cycles(Cyc), Epigenomics(Epi), \\ Genome(Gen), SoyKB(Soy)}\\ \hline
\end{tabular}}
\label{tableZ:benchmark}
\end{table}

For \textbf{DFlow}, \textbf{FaaSFlow}, \textbf{FaasFlowRedis}, and \textbf{CFlow}, we deploy one machine as master node and the global scheduler(GS) is on the master node. Systems utilizes the GS to partition the workflow DAG and assigns the sub-DAG to different machine. We than deploy other 7 machine as worker node to receive the invocation message and execute the function. For \textbf{CFlow}, we deploy the CouchDB on master node for function exchange data. For \textbf{FaaSFlow}, we deploy the CouchDB on master note for inter-node data exchange and deploy Redis on each worker node for intra-node data exchange. The primary distinction between \textbf{FaaSFlowRedis} and \textbf{FaaSFlow} lies in their data exchange mechanisms. FaaSFlowRedis deploys a Redis instance on the master node to facilitate inter-node data exchange, while FaaSFlow employs a CouchDB instance on the master node for the same purpose. 
For \textbf{KNIX}, we deploy the remote Redis on Node 1 and install KNIX on Node2. Additionally, a local Redis instance is deployed on Node 2. We employ the Global Scheduler (GS) from FaaSFlow~\cite{FaaSFlow} to determine the storage type for each function. If a function's storage type is set to DB, KNIX will utilize the remote Redis to store the function's output. Otherwise, KNIX leverages the local Redis instance to save the output of the function. For \textbf{DFlow}, we deploy the data directory service on the master node to store the metadata and deploy the local store on each node.

\subsection{Tail Latency and Throughput} \label{tail-tput} 


In this subsection, we evaluate the impact of throughput changes on the DFlow, CFlow, FaaSFlow, FaaSFlowRedis, KNIX respectively. First, we record the 99\%-ile latency for 6 benchmark where the invocation rate is 6/minute and the network bandwidth is 50MB/s. We run this experiment in open loop. The Figure \ref{fig:AllTailLatencyl} shows the result. In the Figure \ref{fig:AllTailLatencyl}, the x-axis represents the benchmark and the y-axis represents 99\%-ile latency and if one benchmark is timeout, we record its 99\%-ile latency as 60s. 

As shown in the Figure \ref{fig:AllTailLatencyl}, only the Cyc of CFlow is timeout. By exploiting data locality, Both systems except CFlow reduce network bandwidth usage and decrease the 99\%-ile latency. On average, DFlow demonstrates significant improvements in 99\% ile-latency for all benchmark workloads. Specifically, on average, DFlow attains a 52\% reduction in latency compared to CFlow, a 28\% reduction in comparison to FaaSFlow, a 20\% reduction relative to FaaSFlowRedis, and a 36\% reduction when contrasted with KNIX. In CFlow and FaaSFloW, performance deterioration is observed when relying on remote CouchDB for remote data exchange operations, whereas others employ remote Redis for inter-node data exchange under constrained network bandwidth scenarios. This leads to functions competing for limited shared resources, thus causing performance degradation. In contrast, DFlow embraces a dataflow-oriented invocation approach for function execution and leverages a fine-grained data exchange tactic to minimize data movement between functions, consequently reducing tail latency. Additionally, its decentralized, receiver-driven coordination mechanism efficiently optimizes network bandwidth utilization and mitigates contention.




 \begin{figure}[htbp]
     \centering
     \includegraphics[width=\linewidth]{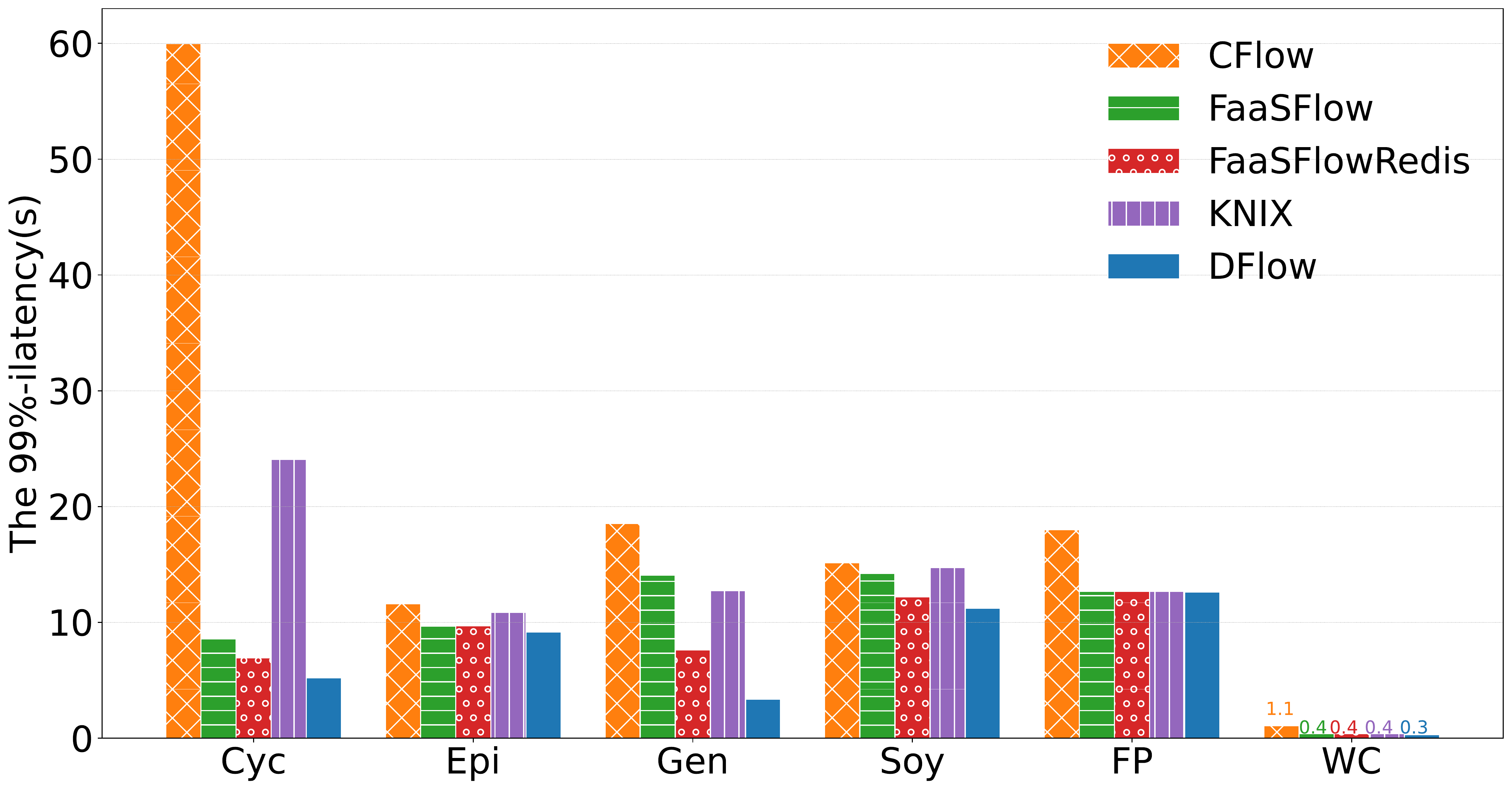}
     \caption{The 99\%-ile latency of CFlow,FaaSFlow and DFlow, when benchmarks running under a 50MB/s bandwidth setup (by sending 6 invocations per minute for each one). The bar reach 60 is timeout. }
     \label{fig:AllTailLatencyl}
 \end{figure}

\begin{figure*}[h]
\centering
\subfigure{\includegraphics[width=40mm]{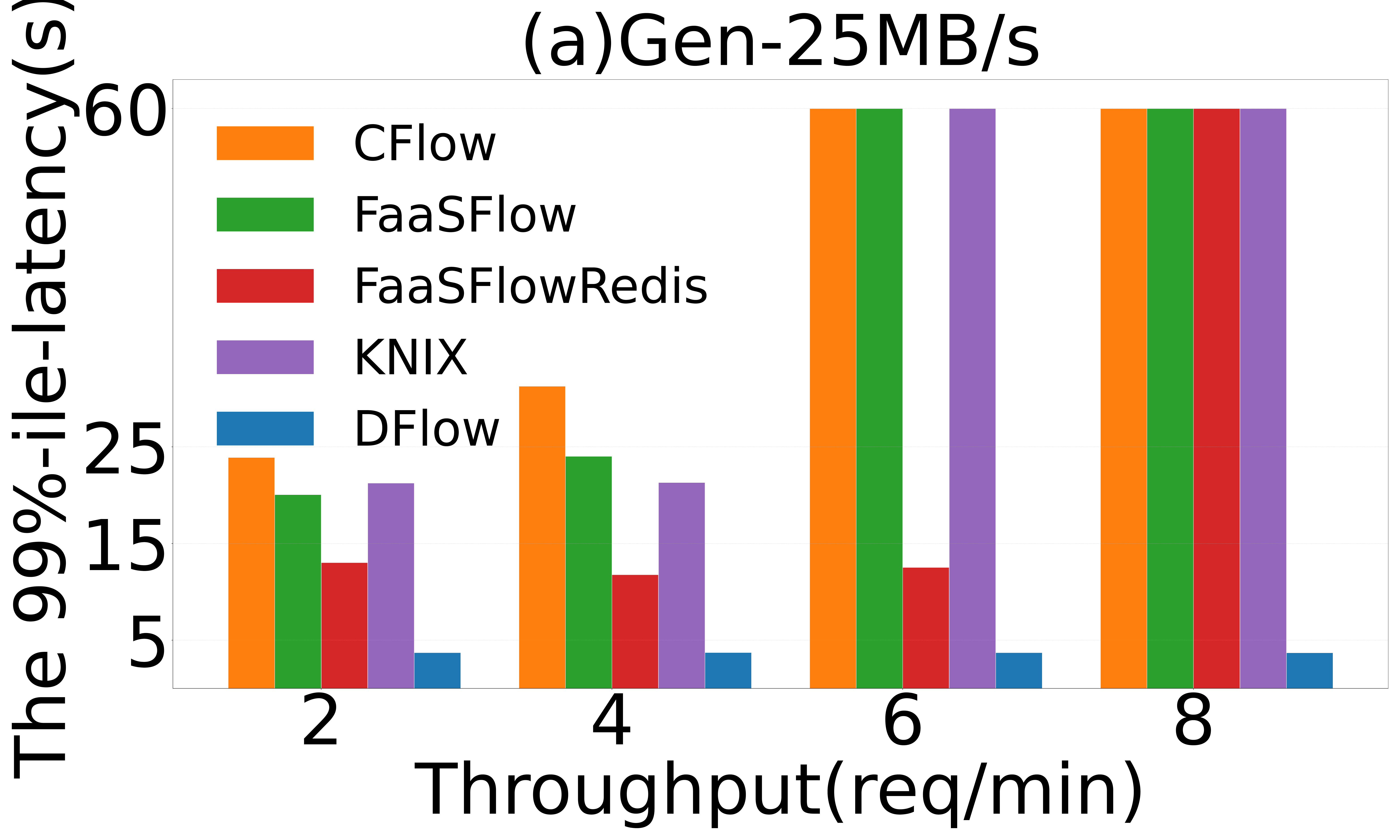}}
\subfigure{\includegraphics[width=40mm]{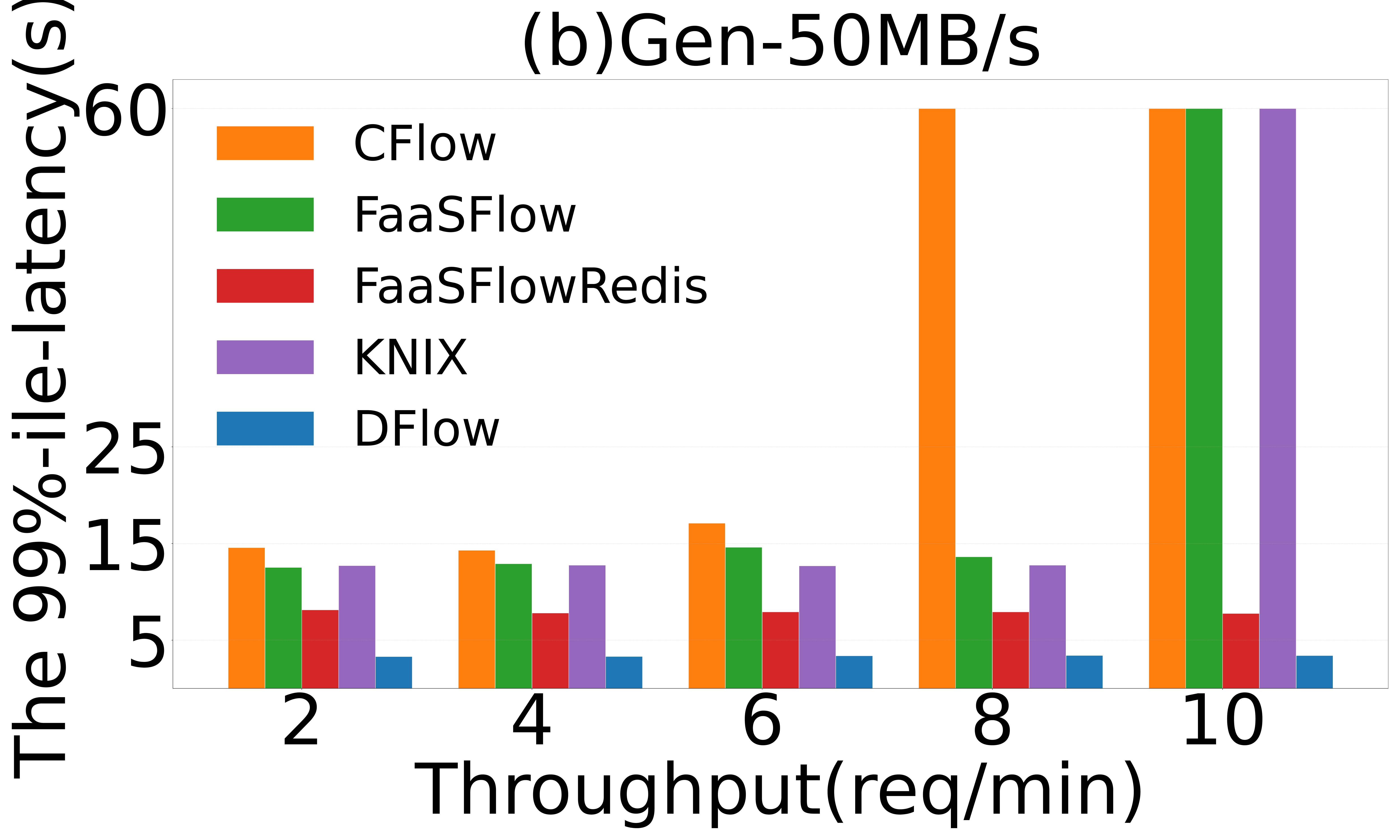}}
\subfigure{\includegraphics[width=40mm]{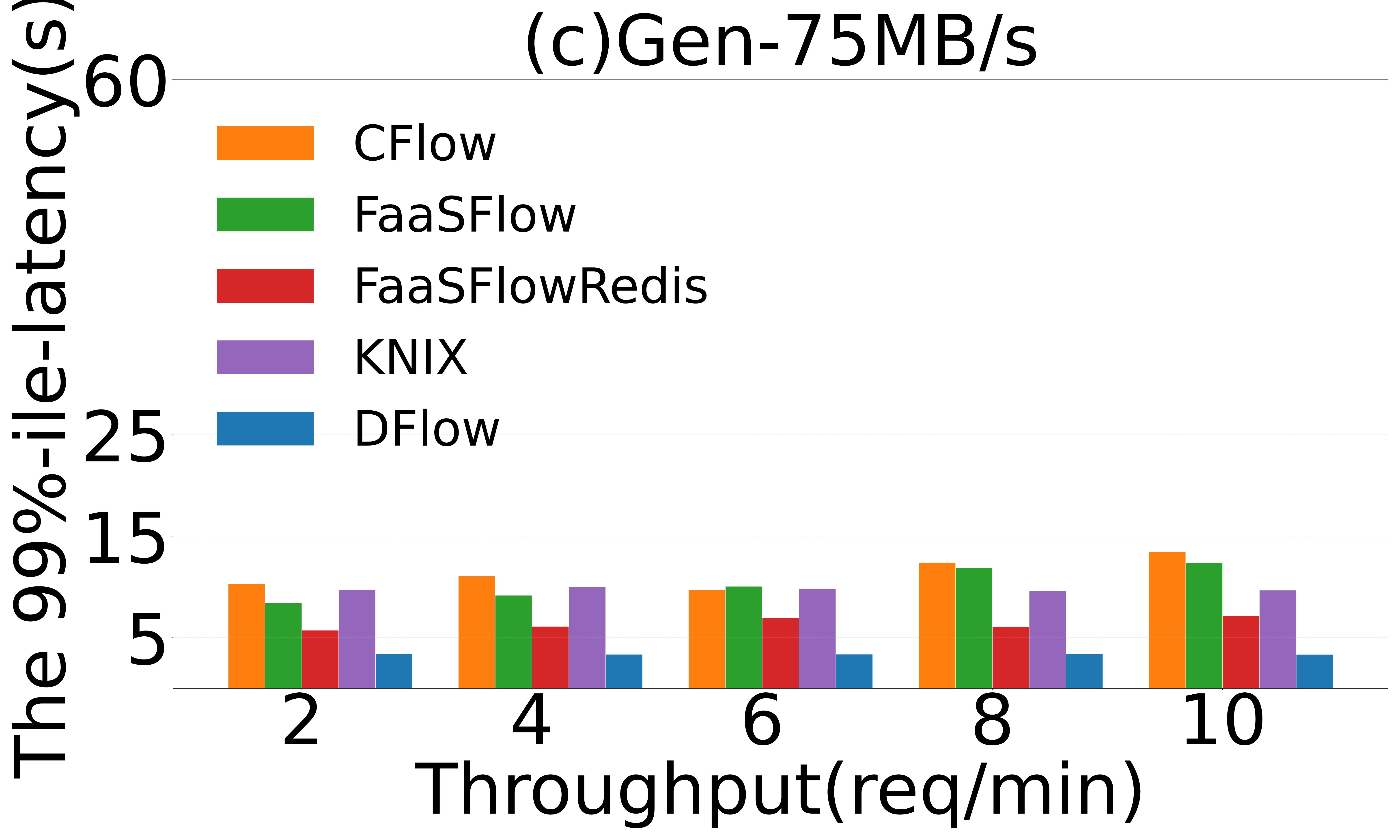}}
\vspace{-2mm}
\subfigure{\includegraphics[width=40mm]{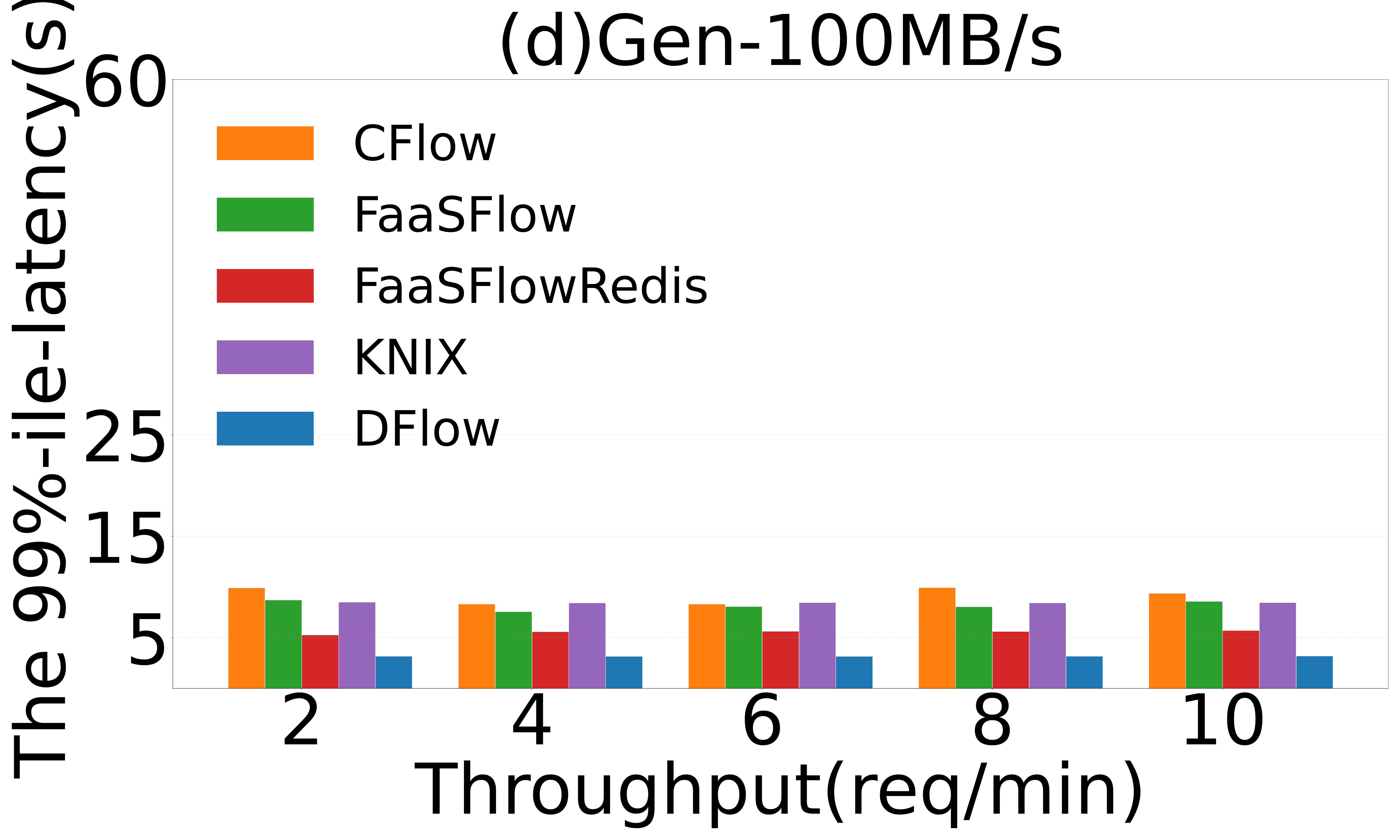}}
\subfigure{\includegraphics[width=40mm]{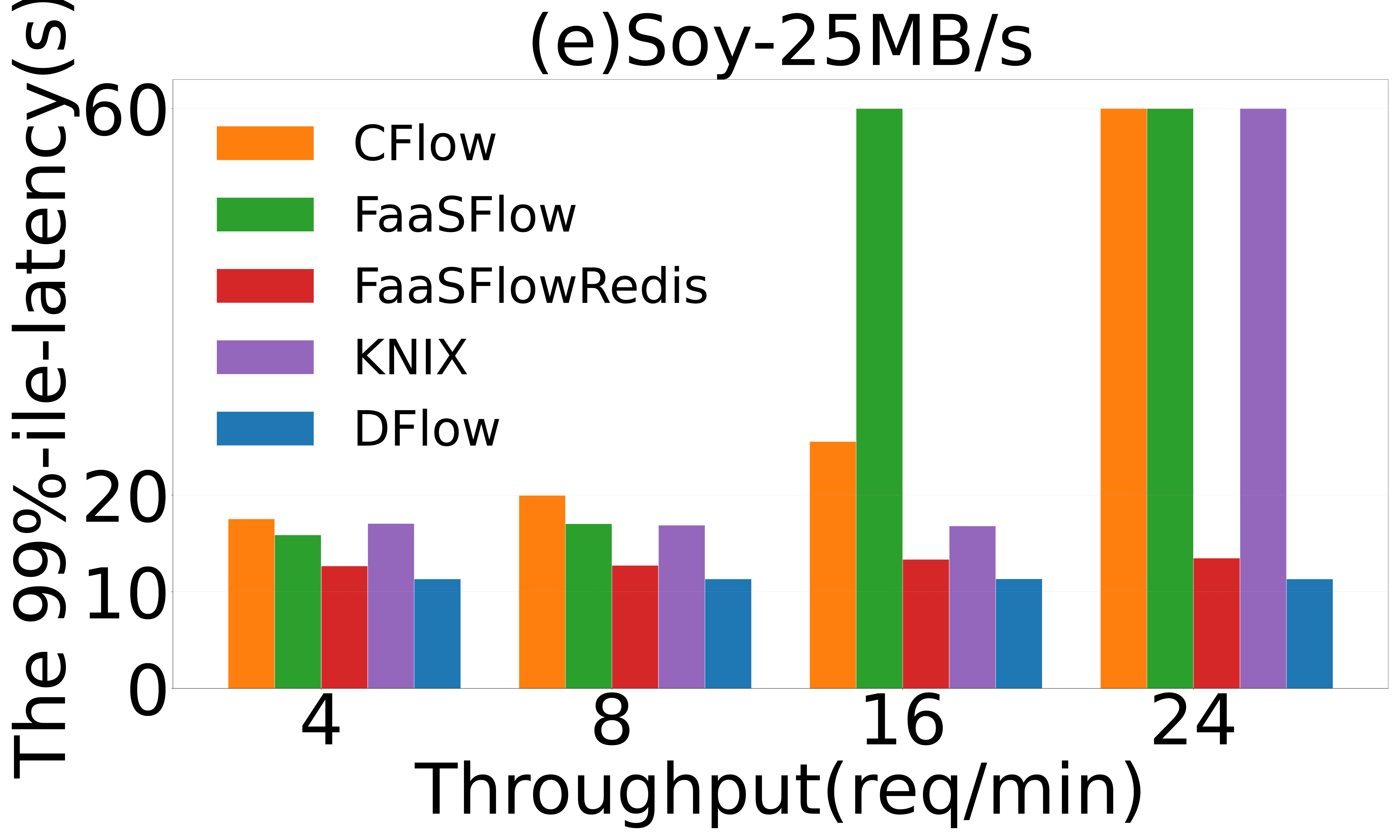}}
\subfigure{\includegraphics[width=40mm]{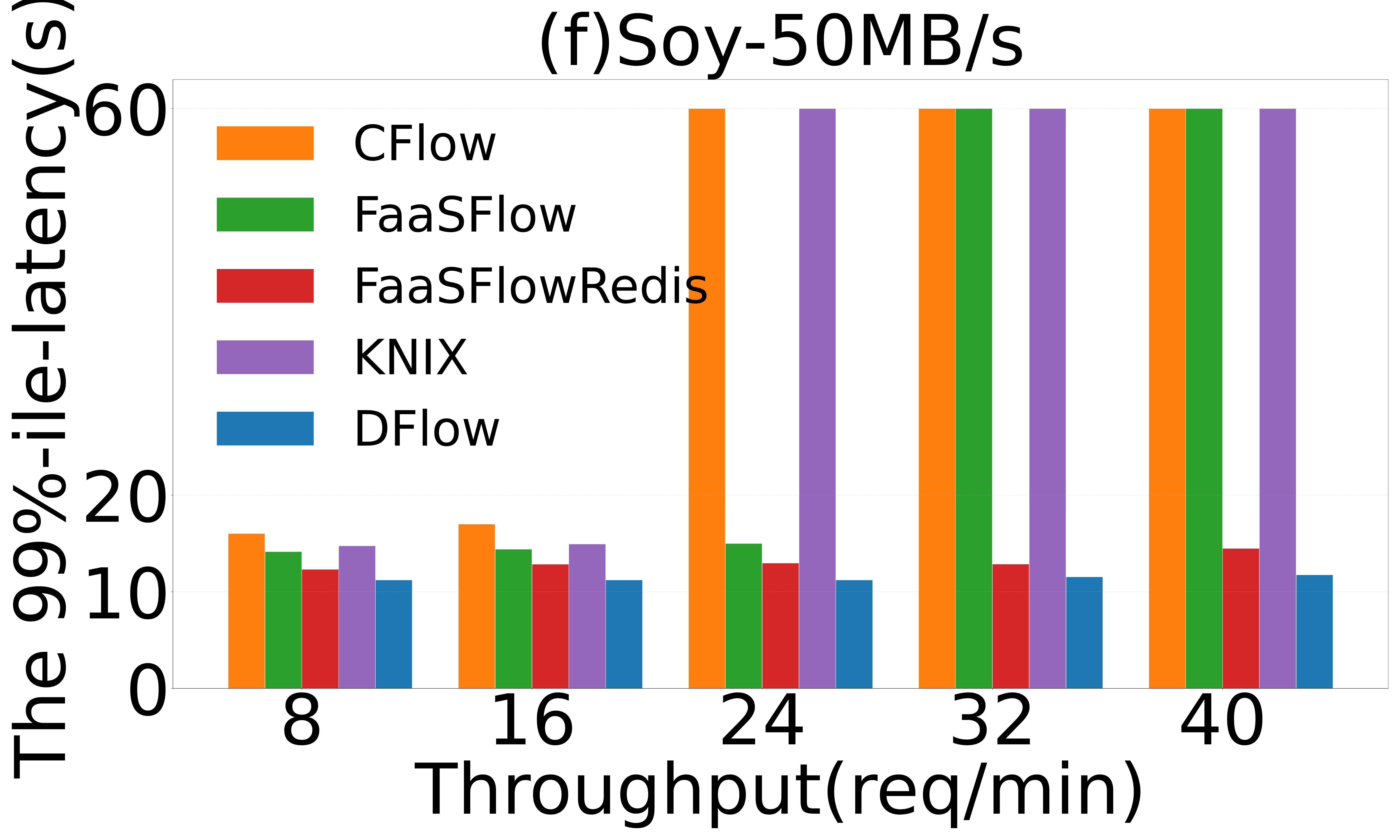}}
\subfigure{\includegraphics[width=40mm]{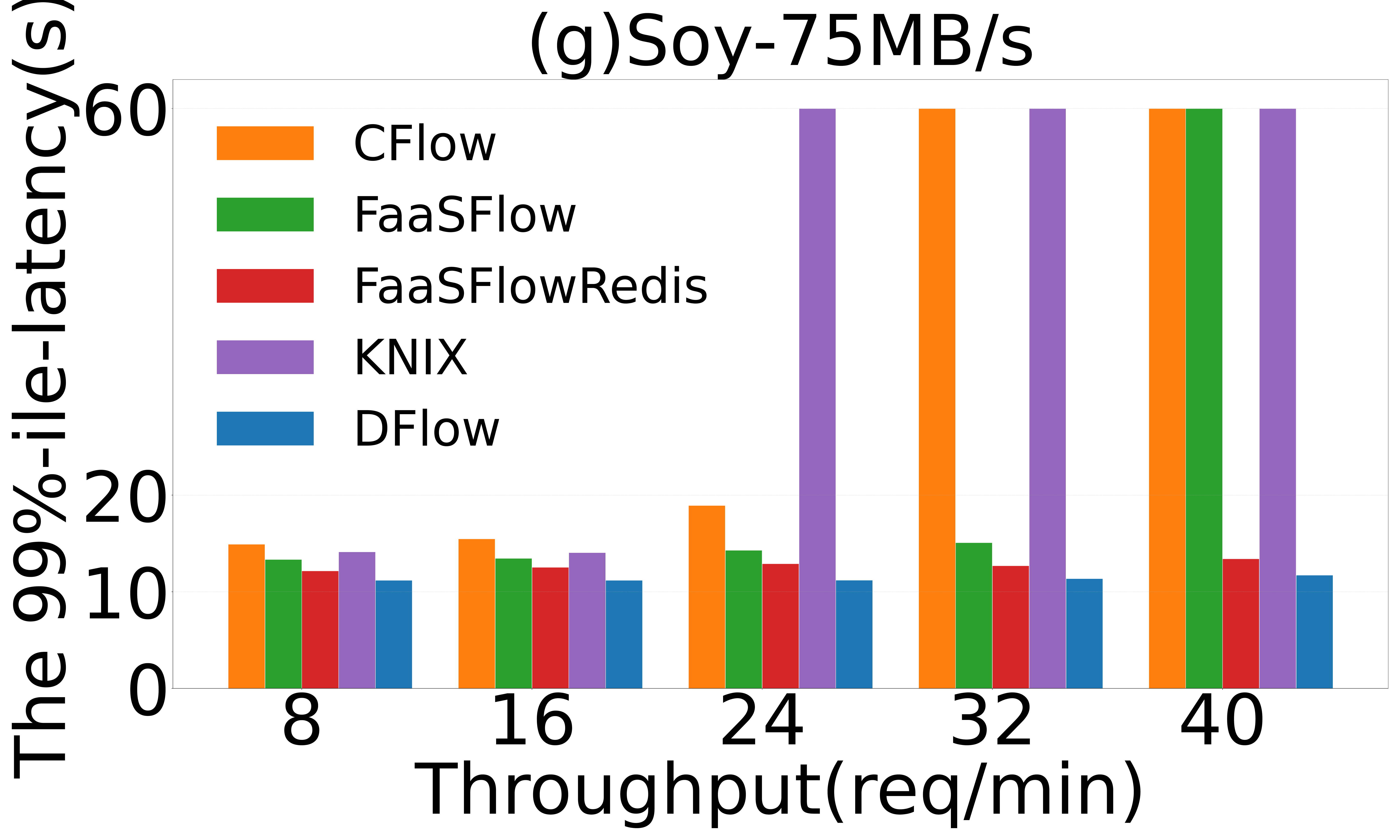}}
\subfigure{\includegraphics[width=40mm]{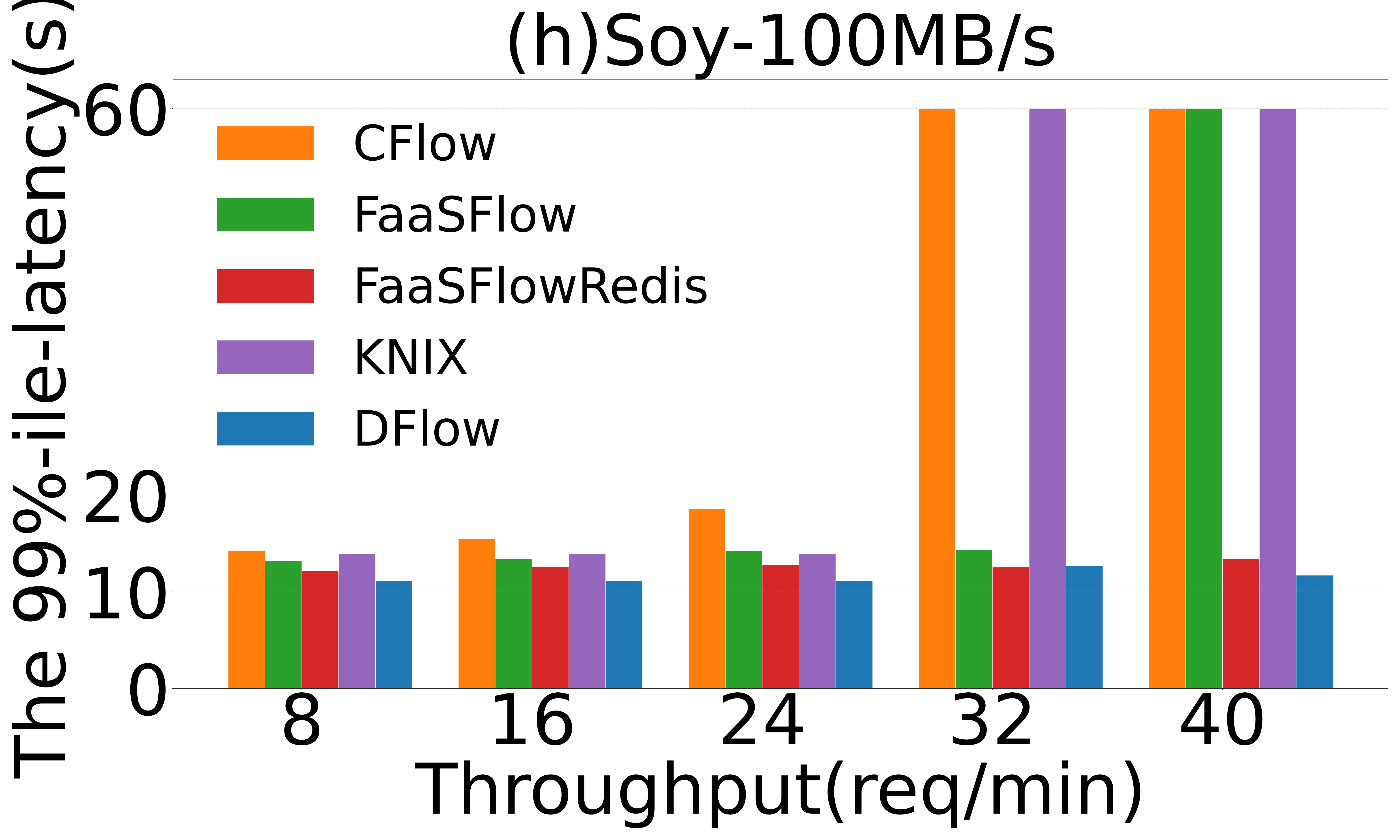}}
\caption{The 99\%-ile latencies under different thourghout for Gen and Soykb when the network bandwidth of the storage node
is configured with 25MB/s, 50MB/s, 75MB/s, and 100MB/s. The bar reach 60 is timeout.}
\label{fig:SoyGen}
\end{figure*}
Subsequently, we evaluate the 99\%-ile latency of CFlow, FaaSFlow, FaaSFlowRedis, KNIX, and DFlow by conducting experiments under varying network bandwidths and invocation rates. We execute these experiments in an open-loop configuration, where the invocation rate corresponds to the throughput.

We use environment introduced in section \ref{benchmarkExp} to run the experiment. For CFlow and FaaSFlow, we employ wondershaper\cite{wondershaper} to set different network bandwidths on the remote CouchDB. In the case of FaaSFlowRedis, and KNIX, we utilize wondershaper [1] to establish varying network bandwidths on the remote Redis instance. For DFlow, considering that the local store on each node transfers data to other local stores, we employ wondershaper to set network bandwidth constraints on each individual local store. Due to space constraints, we focus our evaluation on the Gen and Soykb benchmarks, with other benchmarks exhibiting similar results. The outcomes are illustrated in Figure \ref{fig:SoyGen}.



The Gen benchmark has more than 50 functions and the global scheduler partitions it and schedules it to multiple machine and the data size to exchange is large. As depicted in Figure \ref{fig:SoyGen}, when the network bandwidth is limited and the invocation rate is high, all the baseline systems experience timeouts. Thses can be attributed to their reliance on remote storage for inter-node data transfer, resulting in intense competition among functions for scarce network bandwidth. Conversely, DFlow's decentralized receiver-driven coordinator mechanism proficiently alleviates this contention for shared network resources, 

Under low network bandwidth conditions (25MB/s, 50MB/s), on average, DFlow can achieve a 87\% latency reduction compared to CFlow, an 82\% reduction relative to FaaSFlow, a 65\% reduction in comparison to FaaSFlowRedis, and an 82\% reduction when contrasted with KNIX across all invocation request rates.


when the network bandwidth is high(75MB/s, 100MB/s), all systems can achieve higher throughput and low latency. For instance, at 100MB/s and an invocation rate of 8/min, KNIX is 34\% faster compared to its performance at 50MB/s with the same invocation rate. DFlow consistently demonstrates the best performance across all invocation rates when compared to the other systems under high network bandwidth. Throughout the various invocation request rates, DFlow demonstrates superior performance on average, achieving a 67\% latency reduction in comparison to CFlow, a 64\% reduction with respect to FaaSFlow, a 44\% reduction against FaaSFlowRedis, and a 63\% reduction when pitted against KNIX. 

The enhanced performance of DFlow stems from its dataflow-based invocation pattern, which effectively reduces unnecessary waiting time, combined with the fine-grained data exchange mechanism and efficient decentralized receiver-driven coordination mechanism. These features empower DFlow to effectively minimize network bandwidth contention and fully harness the available bandwidth. As a result, a significant reduction in data exchange between functions is achieved. This, in turn, enables DFlow to attain reduced latency and superior throughput compared to baseline systems.
Likewise, in the Soykb benchmark, DFlow consistently exhibits superior performance on average, achieving a 54\% speed improvement over CFlow, a 41\% improvement compared to FaaSFlow, an 11\% improvement relative to FaaSFlowRedis, and a 51\% improvement when contrasted with KNIX in terms of 99t\%-ile latency across all invocation rates and network bandwidths. The reason DFlow achieves lower latency compared to other systems is analogous to the factors contributing to the Gen benchmark's performance.

In summary, DFlow enhances network bandwidth utilization by up to 4x compared to CFlow and up to 3x compared to FaaSFlow, FaaSFlowRedis, and KNIX. Moreover, it achieves higher throughput and lower 99\% tail latency than its counterparts, including CFlow, FaaSFlow, FaaSFlowRedis, and KNIX.
\subsection{Co-location Interference} \label{co-inter } 
In the serverless cloud platform, it's common that multiple workflow run simultaneously, in which  the multiple workflow will contend for some shared resource, like network bandwidth, CPU and others. Hence, we devise an experimental methodology that consists of two distinct patterns to be executed on CFlow, FaaSFlow, FaaSFlowRedis, KNIX, and DFlow, with the aim of assessing the variations in performance. One pattern is to execute each workflow independently, referred to as a solo-run. Alternatively, workflows can be executed concurrently, known as a co-run. We use the same environment of hardware and software that are introduced in Section \ref{benchmarkExp}. Each client used closed-loop to send invocation. Figure \ref{fig:fllowcorun} shows end-to-end latency result for 6 benchmarks.

\begin{figure*}[h]
\centering
\subfigure{\includegraphics[width=77mm]{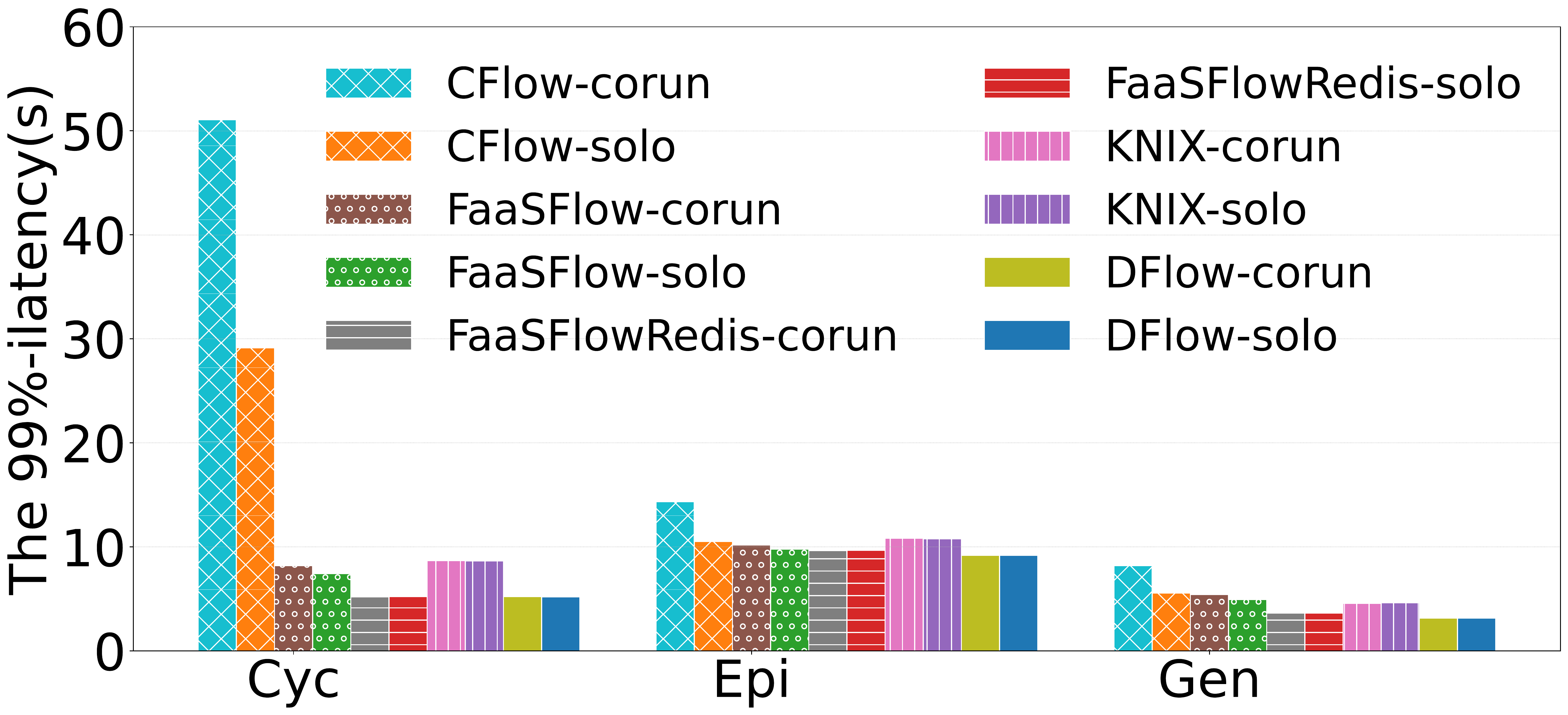}}
\subfigure{\includegraphics[width=77mm]{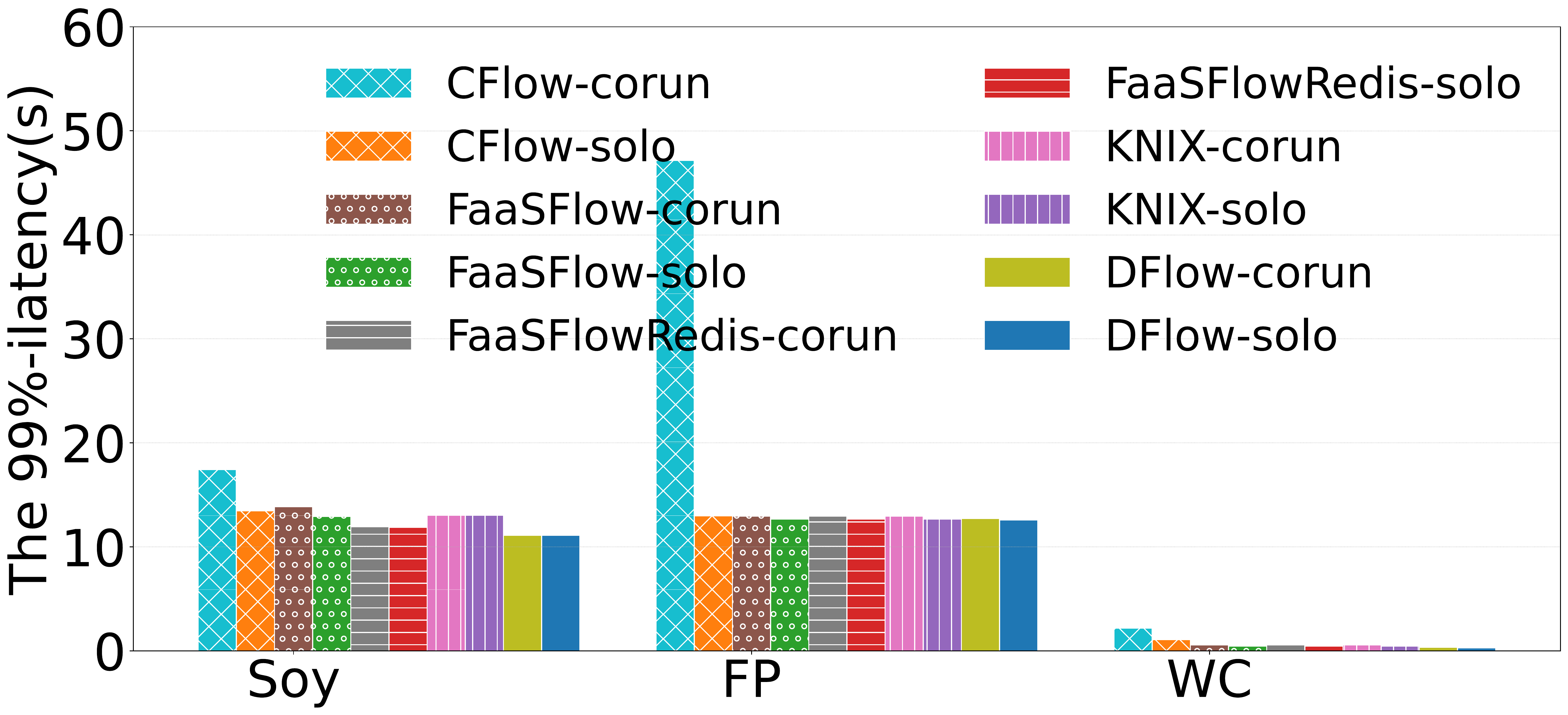}}
\caption{The co-location interference in end-to-end latency for 6 benchmarks when exeucting them in CFlow, FaaSFlow, FaaSFlowRedis, Knix and DFlow}
\label{fig:fllowcorun}
\end{figure*}



As shown in Figure \ref{fig:fllowcorun}, for all systems, excluding CFlow and FaaSFlow, the performance degradation during co-run is minimal compared to their respective solo-run. On average, Corun-CFlow experiences a 40\% performance degradation compared to solo-CFlow, and corun-FaaSFlow degrades by 12\% compared to solo-FaaSFlow. In contrast, other systems only exhibit a 2\% performance decline compared to their solo counterparts. This performance degradation in CFlow and FaaSFlow can be attributed to their use of CouchDB for inter-node data exchange, resulting in high contention for shared resources, including network bandwidth, among others. 

DFlow demonstrates significant performance enhancements, achieving an average improvement of 50\% over CFlow, 22\% over FaaSFlow, 10\% over FaasFlowRedis, and 22\% over KNIX, irrespective of whether executed in a solo-run or co-run scenario. The primary factor contributing to DFlow's superior performance is its utilization of the dataflow-based invocation pattern, which allows for function invocation even while precursor functions are still executing. This approach effectively reduces unnecessary waiting time and enhances overall efficiency. The relatively modest 10\% improvement of DFlow over FaasFlowRedis can be attributed to the close-loop environment, in which an invocation request cannot be sent until the previous request has completed. Consequently, contention for shared resources such as network bandwidth remains low, allowing remote Redis to efficiently perform inter-node data exchange.


\subsection{Cold Start Latency}

 \begin{figure}[t]
     \centering
     \includegraphics[width=\linewidth]{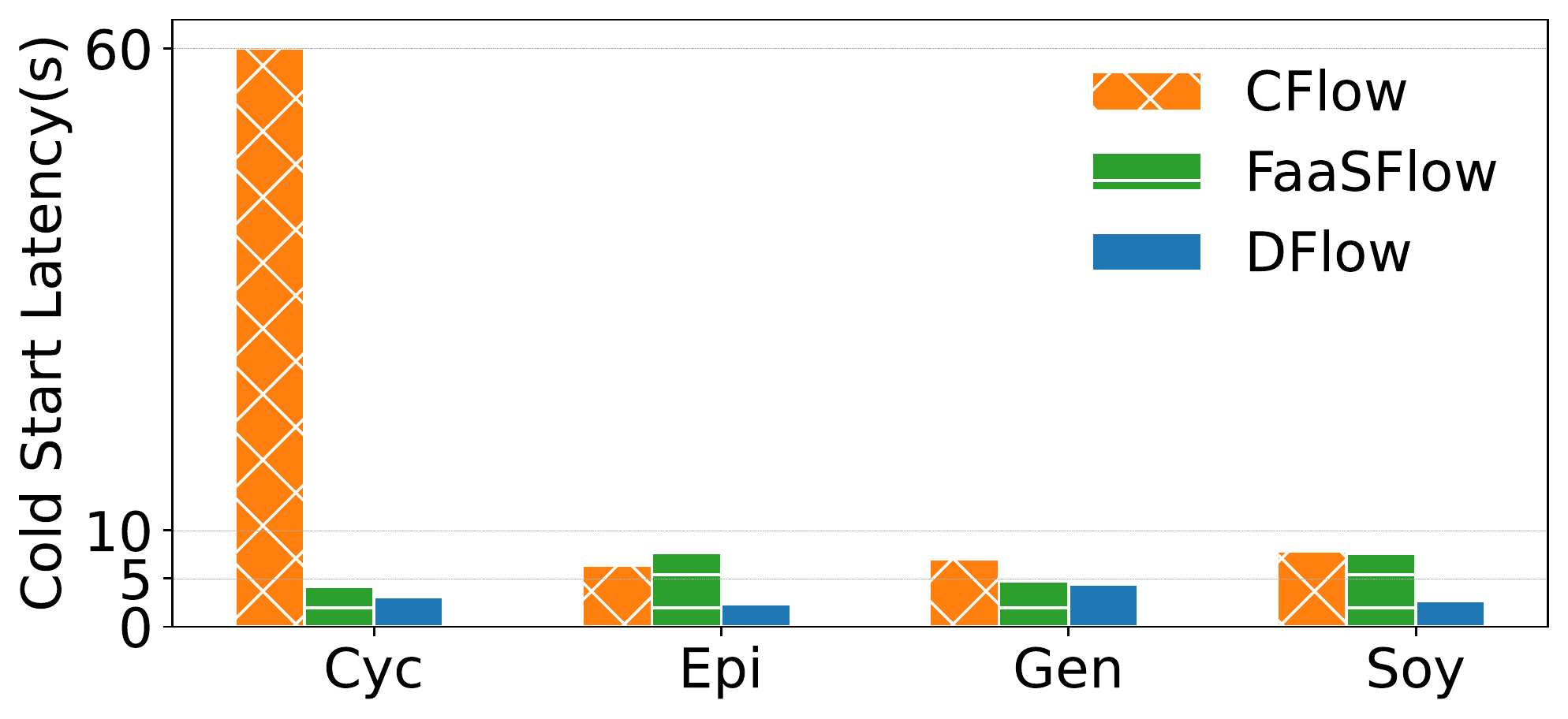}
     \caption{The cold start latency of  CFlow, FaaSFlow, and DFlow  }
     \label{fig:coldstart}
 \end{figure}

Like in Figure~\ref{fig:dataflow-invocation}, DFlow can support function and its precursor functions start to execute simultaneously and their execution time have overlap. Thus, DFlow can reduce the cold start latency. In this evaluation, we focus on comparing DFlow, FaaSFlow, and CFlow. We exclude KNIX from the comparison, as it employs a single container for workflow execution, resulting in a significantly reduced cold start latency. Furthermore, the cold start latency of FaaSFlow is identical to that of FaaSFlowRedis, as the sole distinction between them lies in the remote data exchange approach. FaaSFlow employs remote CouchDB for remote data exchange, whereas FaaSFlowRedis utilizes remote Redis for the same purpose. We carried out experiments in a close-loop setting, involving DFlow, CFlow, and FaaSFlow, by employing the Cyc, Epi, Gen, and Soy benchmarks, respectively, to evaluate their cold start latency.  All benchmarks contain dozens of functions and the functions are distributed on multiple worker node. We measured the difference in end-to-end latency between the first and second runs. The time difference is the cold start latency. The Figure ~\ref{fig:coldstart} shows the result. As shown in Figure ~\ref{fig:coldstart}, DFlow demonstrates a significant reduction in cold start latency, achieving an average improvement of 5.6$\times$ over CFlow and 1.1$\times$ over FaaSFlow.

\subsection{Invocation Pattern Study}
 \begin{figure}[t]
     \centering
     \includegraphics[width=\linewidth]{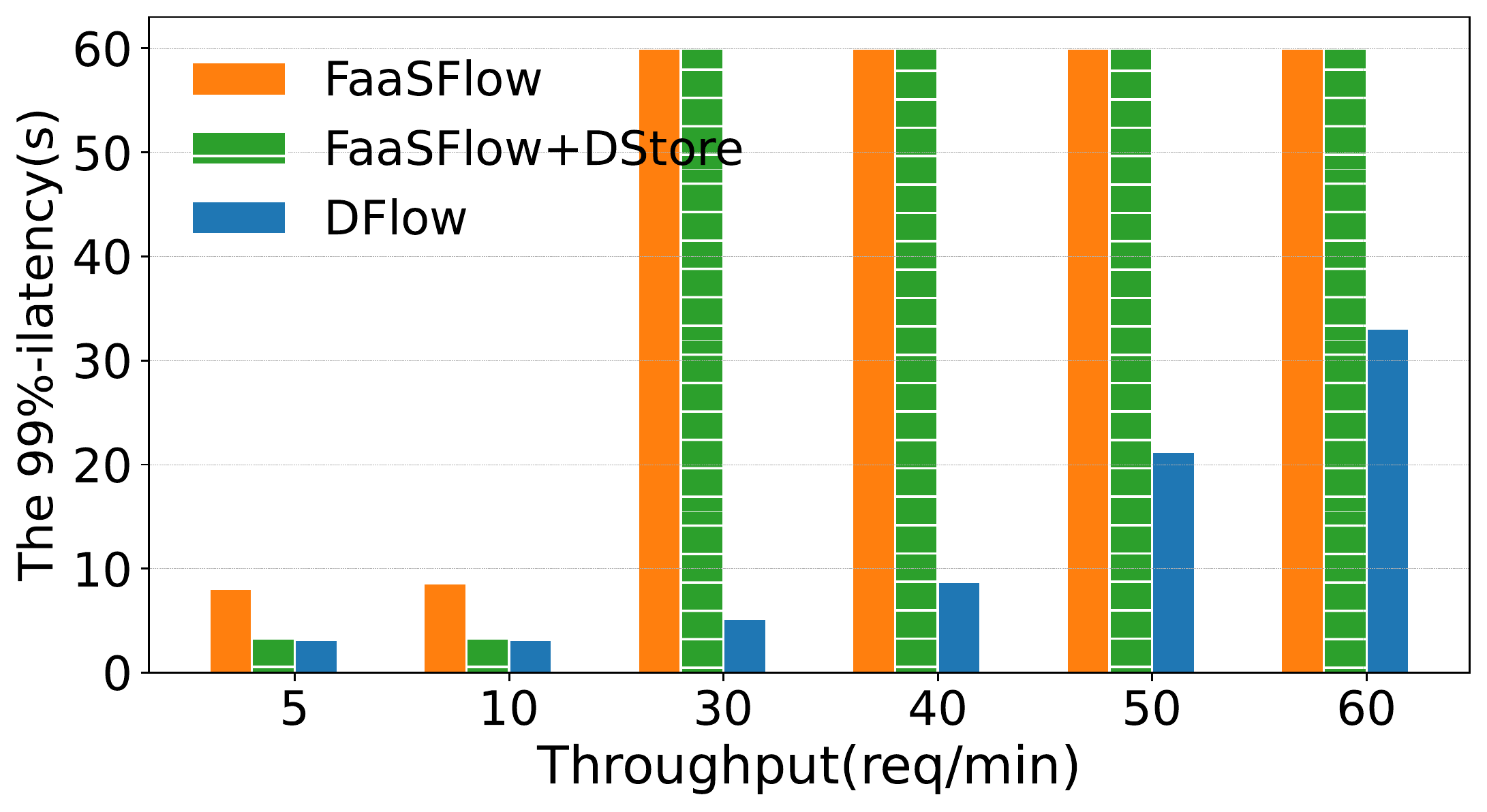}
     \caption{The 99\% tail latency under different throughput for Gen benchmark when the network bandwidth is 100MB/s  }
     \label{fig:invocation}
 \end{figure} 
 
We integrate the DStore into FaasFlow(FaaSFlow+DStore) to study the difference between controlflow-based invocation pattern and dataflow-based invocation pattern. Both of them are decentralized serverless workflow system and use the DStore for efficient data transfter. Due to the space limitation, we use the Gen application in this experiment. Other applications show the similar result.  We evaluate the 99\% latency FaaSFlow, FaaSFlow+DStore and DFlow like Section~\ref{tail-tput} and we set the bandwidth 100MB/s and vary the request rate from 5/min to 60/min. The request rate represents the throughout. 

As show in Figure~\ref{fig:invocation}}, by using the DStore, the FaaSFlow+DStore can achieve 60\% speed compared to FaaSFlow, this is because the DStore can utilize the data locality and make full use of network bandwidth to enable efficient data transfer. When under low request rate, the DFlow can only achieve 5\% speed up on average compared to FaaSFlow+DStore. The modest 5\% improvement can be attributed to the fact that when the request rate is low and contention is relatively minimal, FaaSFlow+DStore effectively utilizes DStore to enable efficient data movement between functions. Consequently, it can achieve performance levels comparable to those of DFlow. But when the request rate is high, both FaaSFlow and  FaaSFlow+DStore are timeout while the DFlow still have good performance. The throughput of DFlow is up to 6 $\times$ over FaaSFlow+Dtore.  This improvement can be ascribed to DFlow's adoption of the dataflow-based invocation pattern, which enables functions to run even while their precursor functions are still executing. This approach effectively reduces unnecessary waiting time and allows the system to handle higher request rates. 

In conclusion, DStore can help FaaSFlow+DStore achieve performance levels similar to DFlow when the request rate is not extremely high. Additionally, the dataflow-based invocation pattern enables DFlow to achieve superior performance under high request rates compared to the control-flow-based invocation pattern.

\section{Related Work}

\textbf{Serverless Workflow System.}

Optimizing the serverless workflow has been a significant research direction, with several notable contributions in recent years~\cite{FaaSFlow, Boki-SOSP21, cloudburst, ORION-OSDI22, Nightcore-ASPLOS21, Faastlane-ATC21, SAND-ATC18, Wukong-SOCC20, Sequoia, Netherite-VLDB22} For instance, FaaSFlow~\cite{FaaSFlow} has proposed a distributed scheduler, consisting of a global scheduler responsible for partitioning the workflow DAG, and a local scheduler that utilizes control flow-based scheduling to invoke functions based on their state. Additionally, FaaSFlow has leveraged FaaStore, a hybrid storage system that uses Redis for local data exchange and CouchDB for remote data exchange, to improve data locality. Similarly, DFlow has used the same global scheduler as FaaSFlow to partition the workflow. A key distinction between FaaSFlow and DFlow lies in their local schedulers. DFlow introduces a dataflow-based local scheduler to support the dataflow-based invocation pattern, whereas FaaSFlow utilizes a control flow-based local scheduler. DFlow utilizes an efficient distributed in-memory store to enable low-latency data exchange between functions. Pheromone~\cite{Pheromone-NSDI23} also utilizes a distributed scheduler to execute a function and leverage a  zero-copy shared-memory object store to support data-centric approach. In Pheromone~\cite{Pheromone-NSDI23}, develops can use data trigger primitives to decide when and how the data passed to the next function. The key distinction between Pheromone and DFlow lies in their distributed schedulers' focus. DFlow's scheduler concentrates on the invocation pattern for function execution, where the output of one function is immediately utilized by the subsequent function. However, DFlow does not have control over when and how data is passed to a function. Cloudburst~\cite{cloudburst} has adopted a distributed in-memory KV store to enable low-latency data exchange, while Boki~\cite{Boki-SOSP21} has employed a shared log to manage the workflow state and support fault tolerance. The primary distinction between DFlow and both BOKI and Cloudburst lies in two aspects: DFlow's implementation of the dataflow-based invocation pattern and its utilization of DStore for fine-grained data exchange.

In the realm of serverless workflow optimization, Nightcore~\cite{Nightcore-ASPLOS21} employs a strategy of scheduling functions from the same workflow onto a single machine to enhance data locality, while leveraging lightweight IPC and efficient IO to achieve low-latency performance. Similarly, Faastlane~\cite{Faastlane-ATC21} adopts a container-based approach to run an entire workflow within a single container and utilizes multi-thread processes to execute functions with minimal data exchange latency. SAND~\cite{SAND-ATC18} leverages different containers to execute different applications, and uses processes to execute functions from these applications, thereby exploiting function locality to enhance performance. Wukong\cite{Wukong-SOCC20} introduces a publisher/subscriber-based serverless parallel framework for task scheduling and partitioning the DAG into sub-DAGs. Sequoia~\cite{Sequoia} presents a novel quality-of-service function scheduling and allocation framework that offers enhanced flexibility and effectiveness in managing serverless function chains. Netherite\cite{Netherite-VLDB22} achieves low latency and high throughput by grouping an application into smaller shards and pipelining the state persistence of each shard. Beldi\cite{Beldi-OSDI20} is a runtime system that supports transactional stateful Serverless Workflows. One key difference between DFlow and these approaches lies in the invocation pattern utilized.

\textbf{Storage System for Serverless.}

Several other approaches exist in the realm of serverless workflow optimization, such as FaasCache~\cite{FaasCache-ASPLOS21}, which uses the Greedy-Dual keep-live policy to reduce cold-start overhead and offers low-latency data exchange. OFC~\cite{OFC-Eurosys21} leverages RAMCloud~\cite{RAMCloud} for caching data, while Infinicache~\cite{InfiniCache} provides fault tolerance and high performance, and low latency by employing erasure coding. Locus~\cite{Locus-NSDI19} proposes a performance model to combine multiple storage types for serverless analytics to deliver high performance and cost efficiency. Pocket~\cite{Pocket} is a storage system that aims to improve data analytics efficiency by decoupling control, metadata, and data management, enabling low-latency and high-throughput performance. Anna~\cite{Anna1,Anna2} is an auto-scaling distributed storage system that offers lower latency and good consistency. HydroCache\cite{ServerlessCasual} offers low latency and achieves Causal Consistency. The crucial distinction between DStore and other approaches lies in the incorporation of fine-grained optimizations within DStore. This design enables efficient data exchange


\textbf{Cold Startup.}

One of the fundamental differences between serverless computing and non-serverless computing is the occurrence of cold start latency for the first function invocation. Numerous works have been proposed to reduce the cold start latency, including Catalyzer~\cite{Yubin}, which employs an innovative OS primitive to reuse the running sandbox. IceBreaker~\cite{Breaker}, which accurately predicts function invocation concurrency and arrival time, and Pagurus~\cite{recycle}, which reuses idle containers from other functions to reduce cold start latency. SAND~\cite{SAND-ATC18} executes all functions of a single application in a container to reduce cold start latency, while Sock~\cite{SockATC18} generalizes Zygote provisioning and introduces a package-aware caching system. These works are complementary to DFlow, which utilizes dataflow-based invocation to create an execution time overlap between functions and their precursor functions, thereby reducing the cold start latency.

\textbf{Misc.}
Hoplite~\cite{Zhuang2021Hoplite} uses the data service directory like DStore to separate the metadata and data and uses the inter-node pipeline to optimize the inter-node data exchange over networks. Ray~\cite{rayOSDI18} also introduces a distributed scheduler and utilizes the global control store to store the metadata.  The primary distinction between DFlow and these systems lies in the fact that DFlow is specifically designed to support dataflow-based invocation patterns in serverless workflows. Consequently, DStore integrates fine-grained optimization and an automatic waking-up/blocking mechanism to facilitate accurate dataflow-based function execution. This focus allows DFlow to effectively handle the unique requirements and challenges of serverless computing environments.

\section{Conclusion} 
In this paper, we propose the DFlow, a decentralized dataflow-based serverless system for serverless workflow. DFlow employs the DScheduler to facilitate dataflow-based invocation patterns, optimizing the execution of functions in serverless workflows. Additionally, DFlow leverages DStore to enable correct and efficient data exchange between functions, ensuring swift data access and transfer. The evaluation of DFlow showcases its ability to significantly reduce end-to-end latency for workflows and minimize cold start latency. Furthermore, DFlow attains high throughput and substantially improves network bandwidth utilization in real-world applications, demonstrating its effectiveness as a solution for serverless workflow systems.

\section*{Acknowledgments}

We thank all the anonymous reviewers for their time and valuable feedback.


\bibliographystyle{unsrt}  
\bibliography{refs}

\end{document}